\documentclass[aip,reprint,floatfix]{revtex4-1}

\draft 

\usepackage{graphicx}
\usepackage{dcolumn}
\usepackage{bm}
\usepackage{braket}
\usepackage{amsmath}
\usepackage{hyperref}
\usepackage{amssymb}
\usepackage{hhline}
\usepackage{xfrac}
\usepackage{color}   %May be necessary if you want to color links
\hypersetup{
    colorlinks=true, %set true if you want colored links
    linkcolor=blue,  %choose some color if you want links to stand out
    citecolor=blue,
    urlcolor=blue
}

\newcommand{\ignore}[2]{\hspace{0in}#2}

\begin{document}
\title{Analysis of trends in experimental observables and reconstruction of the implosion dynamics for direct-drive cryogenic targets on OMEGA} %Title of paper

\author{A. Bose}
\email[]{bosear@umich.edu}
\affiliation{University of Michigan, Ann Arbor MI 48109, USA}
\affiliation{Laboratory for Laser Energetics, University of Rochester, NY 14623, USA}
\affiliation{Department of Physics and Astronomy and/or Mechanical Engineering, University of Rochester, NY 14623, USA}

\author{R. Betti}
\affiliation{Laboratory for Laser Energetics, University of Rochester, NY 14623, USA}
\affiliation{Department of Physics and Astronomy and/or Mechanical Engineering, University of Rochester, NY 14623, USA}

\author{D. Mangino}
%\affiliation{Laboratory for Laser Energetics, University of Rochester, NY 14623, USA}
\affiliation{University of Michigan, Ann Arbor MI 48109, USA}
\affiliation{Department of Physics and Astronomy and/or Mechanical Engineering, University of Rochester, NY 14623, USA}

\author{K. M. Woo}
\affiliation{Laboratory for Laser Energetics, University of Rochester, NY 14623, USA}
\affiliation{Department of Physics and Astronomy and/or Mechanical Engineering, University of Rochester, NY 14623, USA}

\author{D. Patel}
\affiliation{Laboratory for Laser Energetics, University of Rochester, NY 14623, USA}
\affiliation{Department of Physics and Astronomy and/or Mechanical Engineering, University of Rochester, NY 14623, USA}

\author{A. R. Christopherson}
\affiliation{Laboratory for Laser Energetics, University of Rochester, NY 14623, USA}
\affiliation{Department of Physics and Astronomy and/or Mechanical Engineering, University of Rochester, NY 14623, USA}

\author{V. Gopalaswamy}
\affiliation{Laboratory for Laser Energetics, University of Rochester, NY 14623, USA}
\affiliation{Department of Physics and Astronomy and/or Mechanical Engineering, University of Rochester, NY 14623, USA}

\author{O. M. Mannion}
\affiliation{Laboratory for Laser Energetics, University of Rochester, NY 14623, USA}
\affiliation{Department of Physics and Astronomy and/or Mechanical Engineering, University of Rochester, NY 14623, USA}

\author{S. P. Regan}
\affiliation{Laboratory for Laser Energetics, University of Rochester, NY 14623, USA}

\author{V. N. Goncharov}
\affiliation{Laboratory for Laser Energetics, University of Rochester, NY 14623, USA}

\author{D. H. Edgell}
\affiliation{Laboratory for Laser Energetics, University of Rochester, NY 14623, USA}

\author{C. J. Forrest}
\affiliation{Laboratory for Laser Energetics, University of Rochester, NY 14623, USA}

\author{J. A. Frenje}
\affiliation{Massachusetts Institute of Technology, Plasma Science and Fusion Center, Cambridge, Massachusetts 02139, USA}

\author{M. Gatu Johnson}
\affiliation{Massachusetts Institute of Technology, Plasma Science and Fusion Center, Cambridge, Massachusetts 02139, USA}

\author{V. Yu Glebov}
\affiliation{Laboratory for Laser Energetics, University of Rochester, NY 14623, USA}

\author{I. V. Igumenshchev}
\affiliation{Laboratory for Laser Energetics, University of Rochester, NY 14623, USA}

\author{J. P. Knauer}
\affiliation{Laboratory for Laser Energetics, University of Rochester, NY 14623, USA}

\author{F. J. Marshall}
\affiliation{Laboratory for Laser Energetics, University of Rochester, NY 14623, USA}

\author{P. B. Radha}
\affiliation{Laboratory for Laser Energetics, University of Rochester, NY 14623, USA}

\author{R. Shah}
\affiliation{Laboratory for Laser Energetics, University of Rochester, NY 14623, USA}

\author{C. Stoeckl}
\affiliation{Laboratory for Laser Energetics, University of Rochester, NY 14623, USA}

\author{W. Theobald}
\affiliation{Laboratory for Laser Energetics, University of Rochester, NY 14623, USA}

\author{T. C. Sangster}
\affiliation{Laboratory for Laser Energetics, University of Rochester, NY 14623, USA}

\author{D. Shvarts}
%\affiliation{University of Michigan, Ann Arbor MI 48109, USA}
\affiliation{Department of Mechanical Engineering, Ben Gurion University of the Negev, Beer Sheva 84015, Israel}

\author{E. M. Campbell}
\affiliation{Laboratory for Laser Energetics, University of Rochester, NY 14623, USA}

\date{\today}
%
%
%
%%%%%%%%%%%%%%%%%%%%%%%%%%%%%%%%%%%%%%%%%%
%%%%%%%%%%%%%%%%%%%%%%%%%%%%%%%%%%%%%%%%%%
%%				ABSTRACT 
%%%%%%%%%%%%%%%%%%%%%%%%%%%%%%%%%%%%%%%%%%
%%%%%%%%%%%%%%%%%%%%%%%%%%%%%%%%%%%%%%%%%%
%%%%%%%%%%%%%%%%%%%%%%%%%%%%%%%%%%%%%%%%%%
\begin{abstract}
 This paper describes a technique for identifying trends in performance degradation for inertial confinement fusion implosion experiments. It is based on reconstruction of the implosion core with a combination of low- and mid-mode asymmetries. This technique was applied to an ensemble of hydro-equivalent deuterium--tritium implosions on OMEGA that achieved inferred hot-spot pressures $\approx$56$\pm$7 Gbar [S. Regan \textit{et al.}, \href{https://journals.aps.org/prl/abstract/10.1103/PhysRevLett.117.025001}{Phys. Rev. Lett. \textbf{117}, 025001 (2016)}]. All the experimental observables pertaining to the core could be reconstructed simultaneously with the same combination of low and mid modes. This suggests that in addition to low modes, that can cause a degradation of the stagnation pressure, \ignore{ the presence of a systematic mid-mode asymmetry in these implosions. Mid modes reduce the size of the burn volume and cause an overestimation of the ion temperatures measured using neutron diagnostics; the temperatures exhibit large variations ($>$10$\%$) caused by flows introduced by both asymmetries. A thorough analysis of the properties of the x-ray images is provided, this strengthens the argument that these implosions are affected by a combination low and mid modes.}mid modes are present that reduce the size of the neuron and x-ray producing volume. The systematic analysis shows that asymmetries can cause an overestimation of the total areal density in these implosions. It is also found that an improvement in implosion symmetry resulting from correction of either the systematic mid or low modes would result in an increase of the hot-spot pressure from 56 Gbar to $\approx$80 Gbar and could produce a burning plasma when the implosion core is extrapolated to an equivalent 1.9 MJ symmetric direct illumination [A. Bose \textit{et al.}, \href{https://journals.aps.org/pre/abstract/10.1103/PhysRevE.94.011201}{Phys. Rev. E \textbf{94}, 011201(R) (2016)}].
\end{abstract}
% insert suggested PACS numbers in braces on next line
\maketitle %\maketitle must follow title, authors, abstract and \pacs
%
%
%
%%%%%%%%%%%%%%%%%%%%%%%%%%%%%%%%%%%%%%%%%%
%%%%%%%%%%%%%%%%%%%%%%%%%%%%%%%%%%%%%%%%%%
%%				SECTION1: INTRODUCTION 
%%%%%%%%%%%%%%%%%%%%%%%%%%%%%%%%%%%%%%%%%%
%%%%%%%%%%%%%%%%%%%%%%%%%%%%%%%%%%%%%%%%%%
%%%%%%%%%%%%%%%%%%%%%%%%%%%%%%%%%%%%%%%%%%
\section{Introduction}
\label{sec:intro}
 In inertial confinement fusion (ICF),\cite{nuckolls_laser_1972, betti_inertial-confinement_2016} a shell of cryogenic deuterium (D) and tritium (T) filled with DT gas is imploded with either direct laser illumination (direct drive)\cite{bodner_direct-drive_1998, asymth_craxton_direct-drive_2015} or an x-ray bath produced inside a laser-irradiated hohlraum (indirect drive).\cite{lindl_development_1995} Energy from the laser or x ray is absorbed near the outer surface of the shell, causing mass ablation. The shell is imploded to velocities of 300-to-500 km/s to compress the DT gas to high pressures. The shell decelerates during the compression, transferring its kinetic energy to the internal energy of the hot spot. This heats up the low-density (30-to-100 g/cm$^3$) plasma to high central temperatures ($\sim$5 keV) for fusion of D and T nuclei. The hot spot is surrounded and confined by a cold (200-to-500 eV), near-Fermi-degenerate, dense (300-to-1000 g/cm$^3$) fuel layer; the stagnated shell and the hot spot are collectively referred to as the implosion core. 

In this paper we present a systematic analysis of the experimental results for direct-drive implosions and discuss a technique to reconstruct the experimental observables using numerical simulations. The observables are from several cryogenic (DT) implosions on OMEGA\ignore{; these experiments were performed in 2015 and 2016}.\cite{Exp_Regan} The diagnostics of the implosion core include neutron and x-ray detectors. Neutrons are produced from the hot spot by DT fusion reactions; the neutron diagnostics infer the conditions of the hot spot from measurements of the neutron flux, neutron time of flight, and the neutron energy spectrum. High-energy x-ray self-emission from the hot spot, in the 2-to-8 keV range, is imaged using x-ray cameras to infer the shape of the core. 

Observation of repeatable data trends in the direct-drive experiments motivated the development of this analysis technique\ignore{; its applications are eminent}. As the cause of performance degradation for direct-drive implosions is not yet fully identified, we use trends from simulations of the deceleration phase to infer the degradation mechanisms involved. It is known that Rayleigh--Taylor instability (RTI)--induced distortion of the implosion core is a likely cause of degradation; the asymmetries are categorized into low and mid modes, as in Ref. \onlinecite{Bose_physics_2017}, for low modes ($\ell < 6$) the RTI wavelength is longer than the hot-spot radius, whereas for mid modes ($6< \ell < 40$) the asymmetry wavelength is shorter than the hot-spot radius. It was also shown in Ref. \onlinecite{Bose_physics_2017} that the two types of asymmetries have different effects on the neutron-averaged quantities. This paper focuses on trends in the experimental observables arising from asymmetries of the implosion core. The two types of asymmetries are used as the independent basis to approximately reproduce all of the experimental observables. Trends arising from an effective 1-D like degradation, which may be due to shortcomings in the physical models used in hydro codes, are also documented in this paper for future investigations of 1-D degradation.

It is important to emphasize that the experimental observables cannot be explained by using low or mid mode alone, the comprehensive analysis presented here shows that a combination of the two is necessary for the core reconstruction. The exact mode numbers degrading the experiments has not been determined in this paper, other combination of modes could also produce the observables. However, it is shown that in order to reconstruct all the observables simultaneously, the overall balance between the degradation by low modes and the degradation by mid modes must be preserved. 

 The experimental data used in the analysis are summarized in Sec. \ref{sec:Exp_trends}. The reconstruction technique is described in Sec. \ref{sec:Exp_technique}. The trends in the stagnation observables---the inferred pressure, volume, shape, temperature, areal density, neutron burnwith, and bang time---arising from the various degradation mechanisms are also discussed in Sec. \ref{sec:Exp_technique}. Our conclusions along with the future applications for this analysis technique are presented in Sec. \ref{sec:Exp_summary}.
%
%
%%%%%%%%%%%%%%%%%%%%%%%%%%%%%%%%%%%%%%%%%%
%%%%%%%%%%%%%%%%%%%%%%%%%%%%%%%%%%%%%%%%%%
%%				SECTION1: TRENDS IN EXP
%%%%%%%%%%%%%%%%%%%%%%%%%%%%%%%%%%%%%%%%%%
%%%%%%%%%%%%%%%%%%%%%%%%%%%%%%%%%%%%%%%%%%
%%%%%%%%%%%%%%%%%%%%%%%%%%%%%%%%%%%%%%%%%%
\section{Trends in Cryogenic Implosion Experiments}
\label{sec:Exp_trends}
%
%
%
%
%%
%%% %%%%%%                                                FIGURE          %%%%%%%%%%%%%%%
\begin{figure}
\includegraphics[width=80mm]{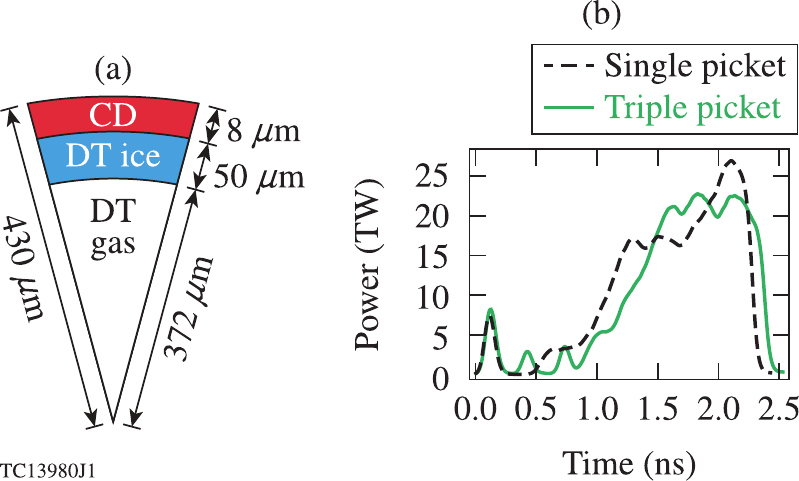}
\caption{\label{fig:Exp_pulse_target} The pulse shapes and targets from the \textit{50 Gbar} implosions.\cite{Exp_Regan}}
\end{figure}
%
%
%
%
%%% %%%%%%                                                TABLE            %%%%%%%%%%%%%%%%
%\clearpage
%\begin{landscape}
\begin{table*}
\caption{\label{tab:Exp_1} This table lists the experimental observable and the corresponding 1-D estimate from simulations [in brackets] for the ensemble of cryogenic implosions on OMEGA which produced $\sim$50 Gbar pressure. In the column showing areal density ($\rho R$) both NTOF and MRS (second) measurements are listed.}
\begin{ruledtabular}
\begin{tabular}{c c c c c c c c c}
%\hline

%& & & & & & &\\

Shot & $Y (\times 10^{13})$ & x-ray $R_{17\%}$ ($\mu$m) & $T_\text{i}$ (keV)\footnote{The ion temperatures were inferred using the instrument response function that was used prior to 2017, currently an updated response function is being investigated, this would result in temperatures that are $\approx$ 300 eV lower than stated, and are within the experimental error.} & $\Delta T_\text{i}$ (keV) & $\rho R$ (mg/cm$^2$) & Burnwidth (ps)& $t_\text{b}-t_\text{b-1D}$ (ps) & $P_\text{inferred}$ (Gbar) \\

 & $\pm 5 \%$ & $\pm 0.5$ $\mu$m & $\pm 0.3$ keV &  & $\pm 31$, $\pm 19$ mg/cm$^2$ & $\pm 6$ ps & $\pm 25$ ps& $\pm 7$ Gbar  \\  \hline
% & & & & & & &\\ \hline %\hhline{|=||=|=|=|=|=|=|=|} %\hhline{=|=|=}
% & & & & & & &\\
\\

78959 &   4.39  & 21.3  & 3.63 & 0.54 & 213, 203 & 71& -16 & 52 \\ 
\vspace{2mm}
         &  [13.8] &[20.9]&  [3.6]&        &  [232]     & [54.1]& & [109] \\
%         & & & & & & &\\ %\hline
%         & & & & & & &\\

78963 & 4.38 & 22.1 & 3.69   &0.88 & 204, 208  & 67   & -20  & 49 \\
\vspace{2mm}
         &[16.3]&[19.8]& [3.74]&       &[242]        &[51.1]&  &[126]  \\  
%         & & & & & & &\\ %\hline
%         & & & & & & &\\

78967 & 3.76  & 21.4  & 3.65& 0.85  & 179, 195 &	64 & -46 & 50 \\
\vspace{2mm}
         & [15.3]&[20.4]&[3.69]&        & [238]     &[51.1]& & [120] \\ 
%         & & & & & & &\\ %\hline
%         & & & & & & &\\

78969 & 4.48  & 21.7   & 3.7   & 0.46 & 204, 197 & 59 & -19  & 55  \\
\vspace{2mm}
         &[14.1] & [21.4] &[3.66]&        &[216]       & [54.7]& & [104] \\ 
%         & & & & & & &\\ %\hline
%         & & & & & & &\\

78971 & 3.77 &	22.1 & 3.69& 1.06 & 220, 208& 72  & -27 &44 \\
\vspace{2mm}
         &[14.4] &[21.4]  &[3.64]&        & [222]     &[52.9]& & [107]  \\ 
%         & & & & & & &\\ %\hline
%        & & & & & & &\\

77064  & 4.21  & 22.0   & 3.32&  0.42& 211, 191& 62  &-26 & 54  \\
\vspace{2mm}
          & [12.5]& [20.4] &[3.48]&      & [219]      & [57.4]& &  [108] \\ 
%          & & & & & & &\\ %\hline
%          & & & & & & &\\

77066  & 4.11  & 21.9  & 3.18& 0.57& 221, 193& 67 & -20 & 56 \\
\vspace{2mm}
          & [16.1]&[21.4]&[3.66]&      &  [228]    &[52.9]& &[112]  \\ 
%          & & & & & & &\\ %\hline
%          & & & & & & &\\

77068  & 5.3  & 22.  & 3.6  &  0.16& 211, 194  & 66 & -31  & 56 \\
\vspace{2mm}
          & [17.]&[22.]&[3.82]&        &[211]       &[61]& &[97]  \\ 
%          & & & & & & &\\ %\hline
%         & & & & & & &\\

77070  & 4.02  & 20.3  &	3.4   & 0.23 & 220, 229  & 70 &-11  & 56  \\
%\vspace{2mm} 
           &[13.3]&[20.4] & [3.55]&        &[239]        &[52.6] & &[114]\\
%           & & & & & & &\\ %\hline
% \hline
\end{tabular} 
\end{ruledtabular}
\end{table*}
%\end{landscape}
%%%%%%%%%%%
%
%
%
%
%
%
%
%
%%% %%%%%%                                                FIGURE          %%%%%%%%%%%%%%%
\begin{figure}
\includegraphics[width=80mm]{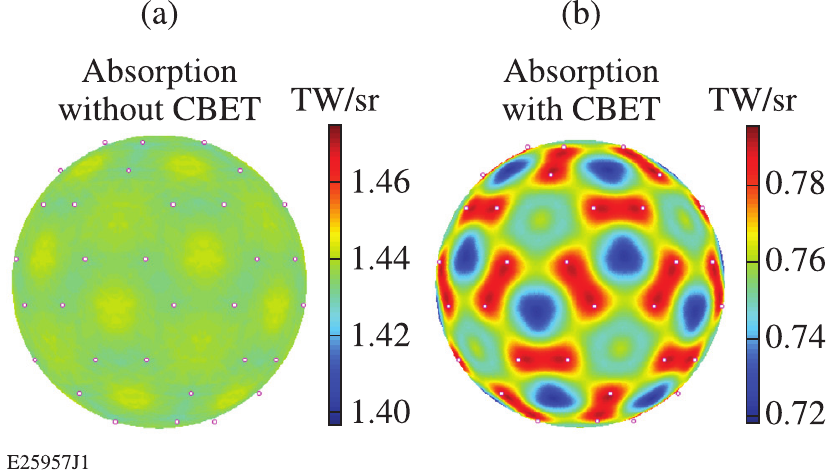}
\caption{\label{fig:Exp_CBET}The laser power absorbed at the target surface is shown for calculations: (a) without considering cross-beam energy transfer (CBET) between the interacting laser beams, and (b) with CBET. [Reprint with permission, LLE Review, Vol. 150\cite{Exp_Edgell}]}
\end{figure}

 It has been shown by Regan \textit{et al.}\cite{Exp_Regan} that direct-drive cryogenic implosions on OMEGA have achieved hot-spot pressures exceeding 50 Gbar---a performance that surpassed all previous implosions on OMEGA. The implosion performance was estimated based on the experimental observables: neutron yield, areal density, ion temperature, hot-spot volume, and neutron burnwidth. The ``\textit{50 Gbar}" implosions used standardized pulse shapes (either a single-picket pulse or a triple-picket pulse) and standardized targets (shown in Fig. \ref{fig:Exp_pulse_target}). The 1-D performance is estimated from simulations using the hydrodynamic code \textit{LILAC}.\cite{Exp_LILAC1} It must be noted that the laser deposition models in \textit{LILAC} were optimized to reproduce in-flight observables like laser-energy deposition and shell trajectory.\cite{Exp_LILAC2, Exp_LILAC3} The estimated implosion adiabat for this design is $\alpha$ $\approx$3.5-to-4 [the adiabat is defined as the ratio of the hydrodynamic pressure $(P)$ and the Fermi pressure of a degenerate electron gas $(P_\text{F})$, at the interface of the hot spot and shell at the time when the laser-driven shocks reach this interface, i.e., $\alpha \equiv P/P_\text{F}$]. This is considered to be a mid-adiabat implosion design, with an adiabat higher than the indirect-drive ``high foot" design.\cite{Exp_hurricane, Exp_doppner, Exp_high_foot} The peak hot-spot pressure in 1-D is estimated to be $\sim$100 Gbar, close to the $\sim$120 Gbar required to demonstrate hydro-equivalent ignition (the hydro-equivalent scaling of the implosion core has been discussed in Refs. \onlinecite{Exp_Regan}, \onlinecite{Bose_RTIscaling_2015} and \onlinecite{Exp_Bose}). Notice that the pressure required for ignition with 1.9 MJ direct illumination is lower than the 350-400 Gbar required for ignition with the indirect drive approach with the same laser energy. This is because for direct drive the conversion efficiency of laser energy to kinetic energy of the imploding shell is much higher, therefore, allowing the implosion of greater DT fuel mass (i.e. larger target radius) which results in longer confinement times ($\tau$). Since the Lawson ignition condition scales as $P_\text{ign} \tau$, the pressure required for ignition ($P_\text{ign}$) is lower with respect to that required for indirect drive. 

Table \ref{tab:Exp_1} lists the performance of several of these \textit{50 Gbar} implosions. The performance parameters are similar for all the shots. The neutron yields are $\sim$4$\times 10^{13}$, at a yield degradation level $Y/Y_\text{1D} \sim 0.3$. Where $Y_\text{1D}$ represents the post-shot 1-D simulation yield, calculated using \textit{LILAC}. The hot-spot radii for all the shots are $\sim$22 $\mu$m; they were estimated using time-resolved x-ray images\cite{Exp_Fred} (discussed in Sec. \ref{sec:Exp_gatedimage}). The ion temperatures ($T_\text{i}\sim 3.5$ keV) are comparable to the temperatures from 1-D simulations, to within 10$\%$ degradation level. The $T_\text{i}$'s were measured using three different detectors---the chemical vapor deposition (CVD) detector \cite{Exp_CVD} and the 12-m and 15-m neutron time-of-flight (nTOF) detectors \cite{Exp_NTOF1, Exp_NTOF2}---positioned along different implosion lines of sight; the minimum temperature is listed in Table \ref{tab:Exp_1}. The variation in $T_\text{i}$ measurement $\Delta T$, which is the difference between the maximum and minimum measured temperatures, is considerable for majority of the shots, ranging between 150 eV-and-1.1 keV. It is observed that the measured areal densities are comparable to the 1-D estimates. The $\rho R$ is measured using the nTOF and magnetic recoil spectrometer (MRS) \cite{Exp_MRS} detectors. The measured burnwidths are slightly longer than the 1-D estimate. The burnwidths are measured using the neutron temporal diagnostic (NTD). \cite{Exp_NTD}

 For direct-drive implosions on OMEGA, it is anticipated that the core is degraded by a combination of low and intermediate modes. Although the origin of the asymmetries is uncertain, low modes can arise from several factors, including long-wavelength target defects, target positioning, laser beam balance and laser beam pointing.\cite{Exp_Hu, Exp_Goncharov, Exp_Igor} In addition, the superposition of all 60 laser beams on OMEGA can produce overlap intensity variations, which is expected to introduce intermediate-mode nonuniformities, similar to the mode $\ell=$10 in 2-D geometry. The cross-beam energy transfer (CBET) calculations by Edgell \textit{et. al},\cite{Exp_Edgell} shown in Fig. \ref{fig:Exp_CBET}, represent the variation in laser-energy absorption at the target surface. When CBET is included, the nonuniformity is higher by 10$\times$. These variations may be associated with the origin of mid-mode asymmetry in direct-drive implosions.
%
%
%
%
%
%
%
%
%
%
%\FloatBarrier
%%%%%%%%%%%%%%%%%%%%%%%%%%%%%%%%%%%%%%%%%%
%%%%%%%%%%%%%%%%%%%%%%%%%%%%%%%%%%%%%%%%%%
%%				SECTION2: Synthetic reconstruction technique
%%%%%%%%%%%%%%%%%%%%%%%%%%%%%%%%%%%%%%%%%%
%%%%%%%%%%%%%%%%%%%%%%%%%%%%%%%%%%%%%%%%%%
\section{The Reconstruction Technique and its Application}
\label{sec:Exp_technique}
%
%
%%% %%%%%%                                                FIGURE          %%%%%%%%%%%%%%%
\begin{figure}
\includegraphics[width=85mm]{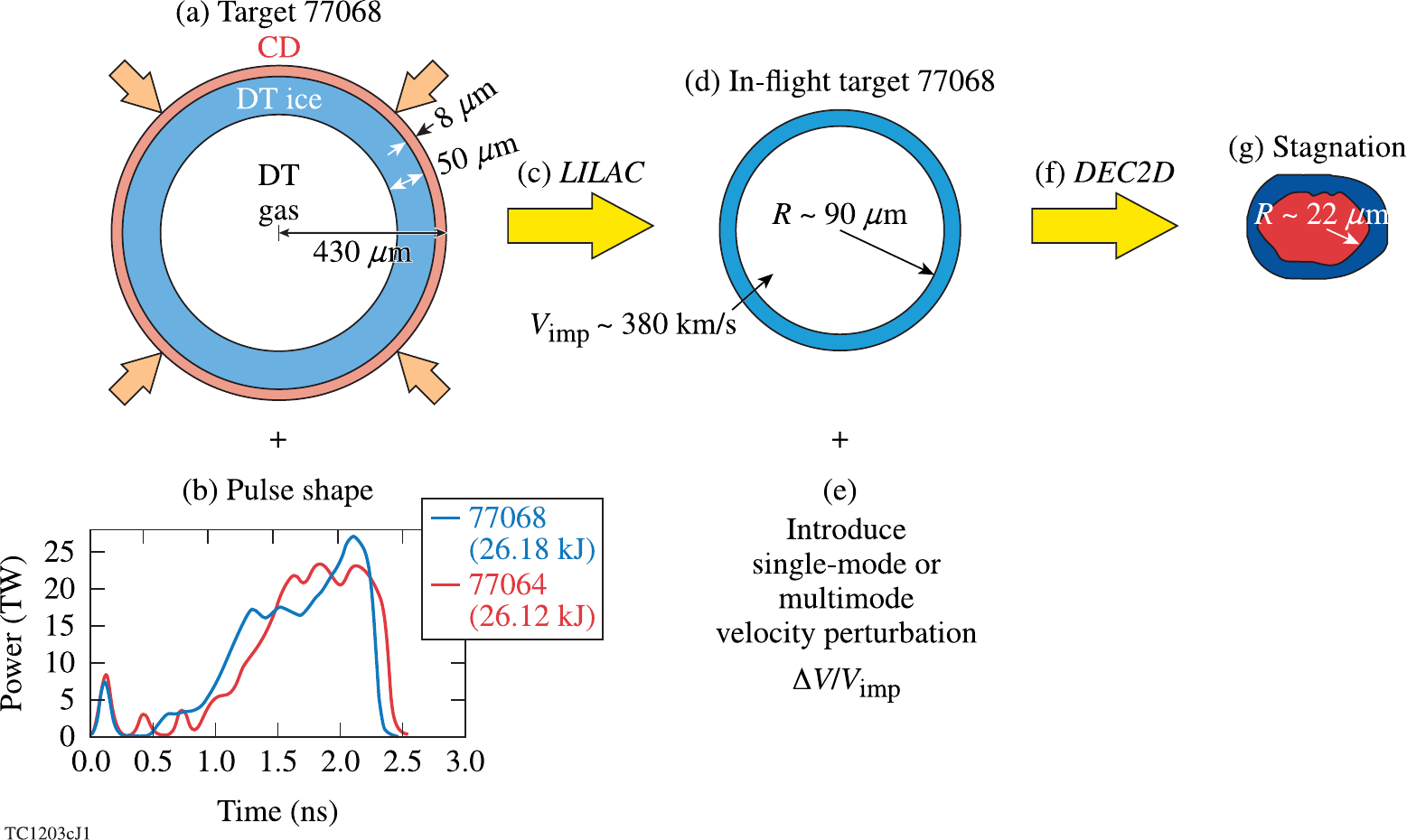}
\caption{\label{fig:Exp_technique} The procedure involved in the reconstruction technique. The (a) target and (b) pulse shape are used as initial conditions for the 1-D hydrodynamic code \textit{LILAC}, which is used to (c) simulate the acceleration phase of implosions. The hydrodynamic profiles from the (d) in-flight target simulation are transferred to \textit{DEC2D}; single- or multimode velocity perturbations are (e) introduced at the inner surface of the shell. (f) The deceleration phase of the implosion is simulated in 2-D; (g) the stagnation parameters are extracted from these simulations.}
\end{figure}
%
%
%
%
%
%%% %%%%%%                                                FIGURE          %%%%%%%%%%%%%%%
\begin{figure}
\includegraphics{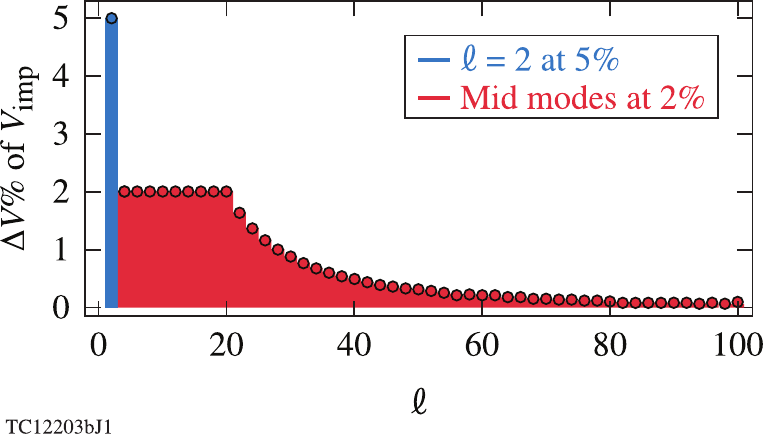}
\caption{\label{fig:Exp_spectrum} The initial velocity perturbation spectrum $\Delta V/V_{\text{imp}} \% (\ell)$ that was used to synthetically reconstruct shot 77068 observables.}
\end{figure}
%
%
%
%
%
%
%
%
%%% %%%%%%                                                FIGURE          %%%%%%%%%%%%%%%
\begin{figure}
\includegraphics[width=80mm]{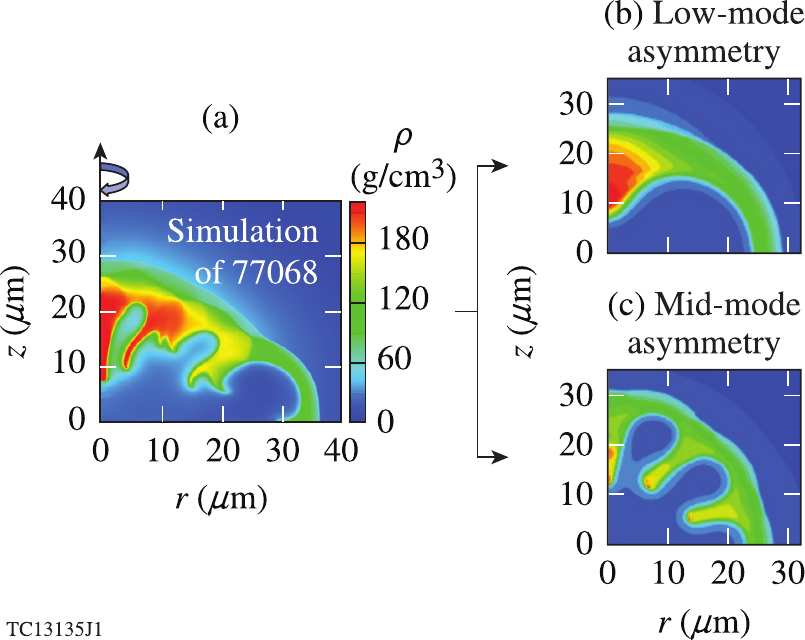}
\caption{\label{fig:Exp_LowMid} Plots illustrating that a combination of low and mid modes were used to reconstruct the core conditions of the shot 77068. The density profiles at time of peak neutron production are shown for (a) the reproduced shot 77068 with $Y/Y_\text{1D} \approx 0.3$, (b) the low-mode $\ell=2$ component at $Y/Y_\text{1D} \approx 0.6$, and (c) an equivalent mid-mode $\ell=10^*$ component at $Y/Y_\text{1D} \approx 0.6$.}
\end{figure}

Unlike the conventional approach that involves full simulations of the implosions including nonuniformities from numerous sources, our technique focuses only on the final phase of an implosion. The final phase consists of the deceleration phase followed by stagnation and disassembly, which are critical in the production of fusion reaction neutrons detected by the nuclear diagnostics, and bremsstrahlung emission detected by the x-ray imaging diagnostics. Performance degradation results from a combination of nonuniformities: they are amplified by the RTI during the acceleration phase and can feed through to the inner surface, where they are further amplified during the deceleration phase by the RTI. 

 The 2-D radiation--hydrodynamic code \textit{DEC2D} is used to simulate the deceleration phase of implosions. The details of the code have been discussed in Ref. \onlinecite{Bose_RTIscaling_2015}. Figure \ref{fig:Exp_technique} provides an outline to our technique, the acceleration phase was simulated using \textit{LILAC},\cite{Exp_LILAC1} it includes the laser drive with models for CBET\cite{Exp_LILAC2} and nonlocal thermal transport.\cite{ Exp_LILAC3} The hydrodynamic profiles at the end of the laser pulse were used as initial conditions for the deceleration-phase simulations in 2-D. Initial perturbations for the deceleration-phase RTI were introduced at the interface of the shell and the hot spot through angular variation of the velocity field.

Here we consider three categories of degradation: low-mode asymmetry, mid-mode asymmetry, and 1-D degradation. The low-mode trends are represented using a mode 2 (``$\ell=2$") and phase reversed mode 2 (``$\ell=2$ \textit{phase reversed}"); the RTI spike axis coincides with the simulation axis of symmetry for the former and they are orthogonal for the latter. The mid-mode trends are represented using a mode 10 (``$\ell=10^*$") and a multimode spectrum referred to as ``\textit{Mid modes}". The $\ell=10^*$ consists of a central mode 10 along with sideband modes 8 and 12 at 20$\%$ of central mode amplitude. The \textit{Mid modes} consist of a spectrum of modes given by $4 \leq \ell \leq 20$ at the same amplitude and a $1/\ell^2$ roll-off spectrum for higher modes $20 \leq \ell \leq 100$, the latter was motivated by the DT ice inner surface roughness spectrum. In simulations, the implosion performance was degraded by increasing the peak amplitude of the velocity perturbation spectrum. The 1-D degradation is incorporated as a degradation in the implosion velocity of the target, i.e., degradation in the initial condition of the deceleration-phase simulations; this has been denoted using ``1-D V$_{\text{imp}}$." The scaling of the implosion observables with $V_{\text{imp}}$ will be shown in the following sections, they are in reasonable agreement with Ref. \onlinecite{Exp_Zhou} which instead uses a set of optimally performing \textit{LILAC} simulations.   

  The single picket pulse shape and target from OMEGA shot 77068 (used in this analysis) are shown in Fig. \ref{fig:Exp_technique} (see blue curve). The analysis technique is very robust and can be applied to any implosion and any scale. The choice of shot 77068 was motivated by the fact that this was the best shot in terms of performance metric $\chi_{\text{no-}\alpha}$\cite{Exp_Regan, Exp_Bose, Betti-alphaheat} and other experimental observables such as yield and areal density. The target was driven with 26.18 kJ of laser energy to an implosion velocity of 380 km/s. The experimental observables, the 1-D simulation parameters, and the reconstructed observables for this shot are shown in Table \ref{tab:Exp_2}. Notice that the experimental observables were reproduced using a combination of the \textit{Mid modes} component (1) and the low mode component (2); a degradation of the simulated 1-D performance with either the low mode or mid modes alone would not produce the estimated results (this can be shown using the last two columns of Table \ref{tab:Exp_2}). The velocity perturbation used for the reconstruction of the shot 77068 is shown in Fig. \ref{fig:Exp_spectrum}; it consists of a combination of low-mode ($\ell=2$) and mid-mode (a spectrum of \textit{Mid modes}) asymmetries. Figure \ref{fig:Exp_LowMid} shows the shape of the hot spot and shell at time of peak neutron production (i.e., bang time $t_\text{b}$); the final shape resembles a combination of a low-mode $\ell=2$, and a dominant mid-mode $\ell=10$. We emphasize that the exact mode numbers degrading the experimental performance cannot be inferred from this analysis technique, and other combinations of modes could also lead to the same reconstructed observables. However, the overall balance between the degradation by low modes and the degradation by mid modes on all of the observables must be preserved. To illustrate this, we also show trends from a different low mode: the $\ell=2$ asymmetry with a reversed phase. Although this mode has a different structure, the resulting trends are the same; for example, see trends in pressure and volume degradation in Figs. \ref{fig:Exp_pressure}, \ref{fig:Exp_volume}, \ref{fig:Exp_temp}, \ref{fig:Exp_rhoR}, and \ref{fig:Exp_BW_BT}. Similarly, the \textit{Mid modes} (of the spectrum in Fig. \ref{fig:Exp_spectrum}) produces very similar degradation trends as the mode $\ell=10^*$.        

 The following sections show the analysis of the \textit{50 Gbar} implosion results using this technique. The effect of low and mid modes on each of the implosion observables is discussed.
%
%
%
%
%
%%%%%%%%%%%%%%
%%%%%%%%%%%%%%
%%% %%%%%%                                                TABLE            %%%%%%%%%%%%%%%%
%%%%%%%%%%%%%%
%%Table1
\begin{table*}
%\begin{center}
\caption{\label{tab:Exp_2}Comparison of measurements with 1-D simulations (using \textit{LILAC} and \textit{DEC2D}) and 2-D simulations (using \textit{DEC2D}).}
\begin{ruledtabular}
%\scalebox{0.85}{
\begin{tabular}{c c c c c c c c}
%\hline
 Observables & Experiment & 1-D simulation & Reconstructed & Mid modes & $\ell=2$ & Mid modes & $\ell=2$ \\
  & shot 77068 &   & shot 77068 & Component (1) & Component (2) & $Y/Y_\text{1D}$$\approx$0.3 & $Y/Y_\text{1D}$$\approx$0.3\\

\hline
%\hhline{|=||=|=|=|=|=|}
\\

Yield& $5.3\times 10^{13}(\pm 5\%)$ & $1.7\times 10^{14}$ & $5.3\times 10^{13}$ & $7.9\times 10^{13}$ & $9.8\times 10^{13}$ & $5.3\times 10^{13}$ & $5.3\times 10^{13}$ \\
$P^*$ (Gbar)& 56($\pm$7) & 97 & 57 & 77 & 73 & 66 & 50\\
$T_{\text{i}}$ (keV)& 3.6($\pm 0.3$) & 3.82 & 3.7 & 3.78 & 3.71 & 3.64 &  3.42  \\
$R_{\text{hs}}$ ($\mu$m)& 22($\pm1$) & 22 & 22 & 20.9 & 23.4 & 21 & 25.3\\
$\tau$ (ps)& 66($\pm 6$) & 61 & 54 & 55 & 56 & 53 & 59\\
$\rho R$ (g/cm$^2$)& 0.194($\pm0.018$) & 0.211 & 0.194 & 0.222 & 0.193 & 0.211 & 0.180 \\
%\hline
\end{tabular}%}
%\end{center}
\end{ruledtabular}
\end{table*}
%%%%%%%%%%%%%%
%%%%%%%%%%%%%%
%
%
%
%
%
%
%\FloatBarrier
%%%%%%%%%%%%%%%%%%%%%%%%%%%%%%%%%%%%%%%%%%
%%%%%%%%%%%%%%%%%%%%%%%%%%%%%%%%%%%%%%%%%%
%%				SECTION3: Pressure
%%%%%%%%%%%%%%%%%%%%%%%%%%%%%%%%%%%%%%%%%%
%%%%%%%%%%%%%%%%%%%%%%%%%%%%%%%%%%%%%%%%%%
%%%%%%%%%%%%%%%%%%%%%%%%%%%%%%%%%%%%%%%%%%
\subsection{Inferred Hot-Spot Pressure}
\label{sec:Exp_pressure}
%
%
%
%
%%% %%%%%%                                                FIGURE          %%%%%%%%%%%%%%%
\begin{figure}
\includegraphics[width=80mm]{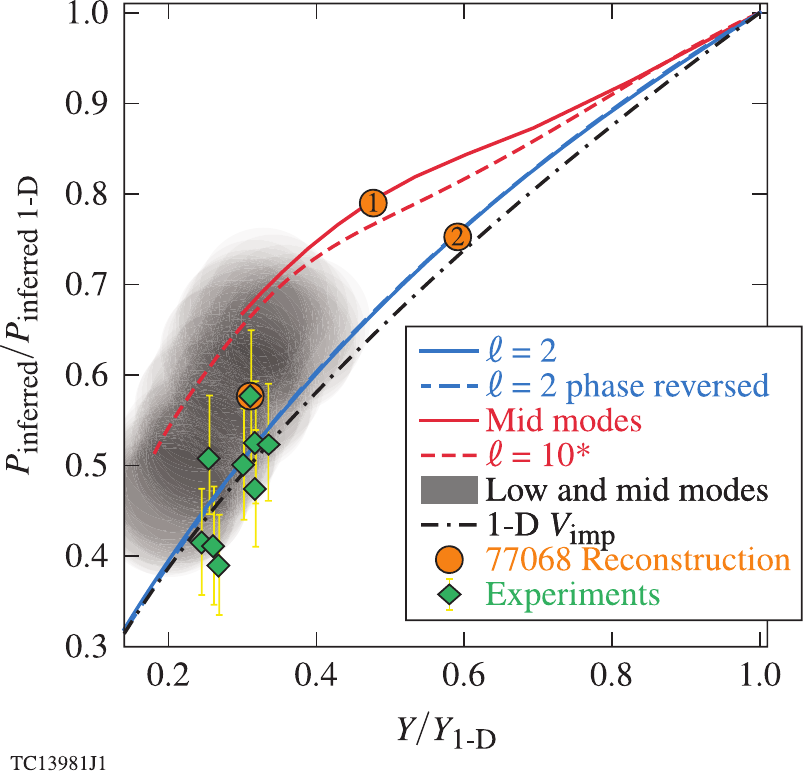}
\caption{\label{fig:Exp_pressure} The degradation in inferred hot-spot pressure $P_\text{inferred}$, normalized with 1-D pressure ($P_\text{inferred-1D}$), versus degradation in yield ($Y/Y_\text{1D}$). This pressure is computed using Eq. \ref{eqn:Exp_pressure} and the x-ray volume. The \textit{50 Gbar} shots in Table \ref{tab:Exp_1} are shown in green. The reconstructed shot 77068 is shown in orange (overlapping the experimentally inferred pressure for shot 77068), with points (1) and (2) representing the degradation caused separately by the mid-mode and low-mode components. The gray-shaded region represents an ensemble of simulations using different amplitude combination of $\ell=2$ and \textit{Mid modes}; it is observed that these reproduce the experiments approximately.}
\end{figure}
The hot-spot pressure is not directly measurable but it is inferred from other experimental observables using \cite{Exp_SC}
%
%
%
%-----------
\begin{eqnarray}
\label{eqn:Exp_pressure}
\frac{P_\text{inferred}}{P_\text{inferred 1D}} =  \hspace{55mm} \nonumber
\\
\sqrt{ \left( \frac{Y}{Y_\text{1D}} \right)  \left( \frac{V}{V_\text{1D}} \right)^{-1}   \left[ \frac{\left( \left< \sigma v  \right>/T_\text{i}^2 \right)}{\left( \left< \sigma v  \right>/T_\text{i}^2 \right)_\text{1D}} \right]^{-1}  \left( \frac{\tau}{\tau_\text{1D}} \right)^{-1}  } \hspace{5mm}, \nonumber
\\
\end{eqnarray}
%-----------
%
%
where $Y$ is the implosion yield obtained from experiments or simulations and is normalized with the 1-D yield ($Y_\text{1D}$) from simulations. The $V/V_\text{1D}$ is the normalized volume of the hot spot, calculated from the x-ray images of experiments or simulations. The fusion reactivity is a function of temperature only,\cite{Exp_Bosch} $\left< \sigma v \right>/T_\text{i}^2 \sim T_\text{i}^\sigma$, with $\sigma \approx 1$-to-$2$ for the temperature range of interest to ICF. The neutron burnwidth $\tau$ is the full width at half maximum of the neutron rate. The degradation trends for each of these observables will be shown in the following sections. 

 The degradation in pressure corresponding to a given degradation in yield is shown in Fig. \ref{fig:Exp_pressure}. The degradation in inferred pressure is an outcome of the degradation in all of the measurable parameters shown in Eq. (\ref{eqn:Exp_pressure}). For any yield degradation level, the low modes (in blue) result in a greater degradation of the hot-spot pressure as compared to mid modes (in red). The $\ell=2$ and $\ell=2$ \textit{phase reversed} produce nearly identical pressure degradation curves; also the $\ell=10^*$ and \textit{Mid modes} produce similar curves. This is because for implosions with mid-mode asymmetries the hot-spot volume is smaller as a result of cooling by penetration of the RTI spikes, but for low modes the volume is larger (see Sec. \ref{sec:Exp_gatedimage}). The gray-shaded region represents an ensemble of simulations using different amplitude combinations of $\ell=2$ and \textit{Mid modes}; with the $\ell=2$ amplitude varying between $4 \%$ and $7\%$ of $V_\text{imp}$ and the \textit{Mid modes} amplitude varying between $2\%$ and $4\%$ of $V_\text{imp}$. The initial velocity perturbation spectrum of Fig. \ref{fig:Exp_spectrum} could be used to reproduce the experimental pressure for shot 77068. The dashed black line in Fig. \ref{fig:Exp_pressure} shows the 1-D pressure scaling with implosion velocity; it follows $P_\text{inferred} \sim V_\text{imp}^{3.72}$. The corresponding yield scaling with implosion velocity follows $Y \sim V_\text{imp}^{6.26}$. The implosion velocity degradation is a simplistic method to model the degradation in implosion convergence; it is useful only for comparison of trends. In experiments, degradation in implosion convergence can be caused by the following: very short scale nonuniformities arising from laser imprinting or reduced laser-to-capsule drive with respect to simulation, and preheating caused by super-thermal electrons (which decrease the implosion convergence by increasing the implosion adiabat $\alpha$). 

 Notice that in Fig. \ref{fig:Exp_pressure} the pressure degradation curve for the 1-D V$_\text{imp}$ coincides with the low-mode curves ($\ell=2$ and $\ell=2$ \textit{phase reversed}), but is different from the mid mode curves ($\ell=10^*$ and \textit{Mid modes}). This can be explained based on Ref. \onlinecite{Bose_physics_2017}. It is so because, firstly, the hot-spot is not isobaric for implosions with mid-mode asymmetries, and, secondly, the inferred pressure for mid modes is the average pressure of the x-ray--producing region of the hot spot. The x-ray--producing volume, however larger than the neutron-producing volume, is still smaller than the total hot-spot volume including the bubbles (i.e., $V_{\left< \text{hs} \right>}$ of Ref. \onlinecite{Bose_physics_2017}). As a result, the inferred pressure for implosions with mid-mode asymmetry using the x-ray volume is higher than the average hot-spot pressure. However, for the low mode asymmetry or 1-D V$_\text{imp}$ degradation curves (Fig. \ref{fig:Exp_pressure}) the hot spot is approximately isobaric and the neutron and x-ray volumes are comparable to the total hot-spot volume ($V_{\left< \text{hs} \right>}$, see Fig. 7 of Ref. \onlinecite{Bose_physics_2017}), therefore the inferred pressure are similar. If the neutron producing volume is used instead of the x-ray volume, the inferred pressure for mid-modes would be similar to the clean (1-D) value --irrespective of the yield, also shown in Fig. 7 of Ref. \onlinecite{Bose_physics_2017}. In summary, the inferred pressure for implosions with mid mode asymmetry is higher than that of low modes at the same yield degradation level, this results from a non isobaric hot spot and a smaller hot-spot volume for the former. 

%
%\FloatBarrier
%%%%%%%%%%%%%%%%%%%%%%%%%%%%%%%%%%%%%%%%%%
%%%%%%%%%%%%%%%%%%%%%%%%%%%%%%%%%%%%%%%%%%
%%				SECTION3: SIZE
%%%%%%%%%%%%%%%%%%%%%%%%%%%%%%%%%%%%%%%%%%
%%%%%%%%%%%%%%%%%%%%%%%%%%%%%%%%%%%%%%%%%%
\subsection{Estimation of the Hot-Spot Size: Using Time-Gated Self-Emission Images}
\label{sec:Exp_gatedimage}
\begin{figure}
\includegraphics[width=80mm]{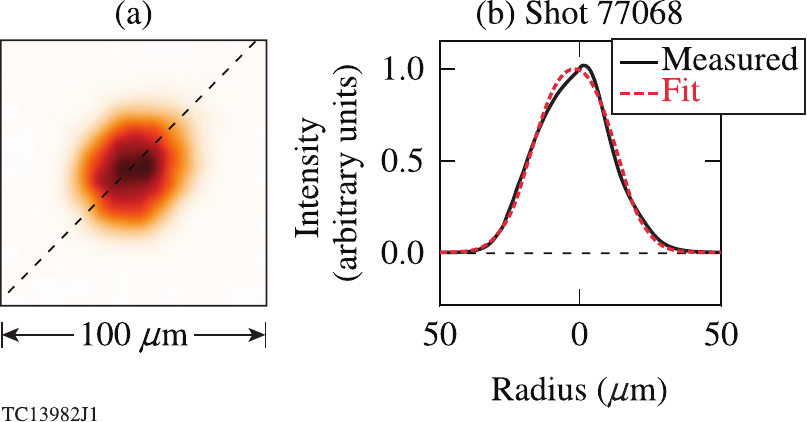}
\caption{\label{fig:Exp_kbframed} (a) An x-ray image of the hot spot at stagnation for shot 77068, obtained using a time-resolved Kirkpatrick--Baez (KB) framed camera with a 4-to-8 keV photon energy range and an $\sim$ 6 $\mu$m spatial resolution.\cite{Exp_Fred} The measured and fit x-ray profiles along the dashed line are shown in (b).}
\end{figure}

 Time-resolved images of the core x-ray self-emission, as shown in Fig. \ref{fig:Exp_kbframed}, have been used to estimate the hot-spot volume.\cite{Exp_Fred} Here $R_{17}$ is the radius at 17$\%$ of peak intensity and $V_\text{x-ray}/V_\text{x-ray-1D} = (R_{17}/R_{17-\text{1D}})^3$.
%
%
%%% %%%%%%                                                FIGURE          %%%%%%%%%%%%%%%
%
%
%
%
%
%
%
%%% %%%%%%                                                FIGURE          %%%%%%%%%%%%%%%
\begin{figure}
\includegraphics[width=80mm]{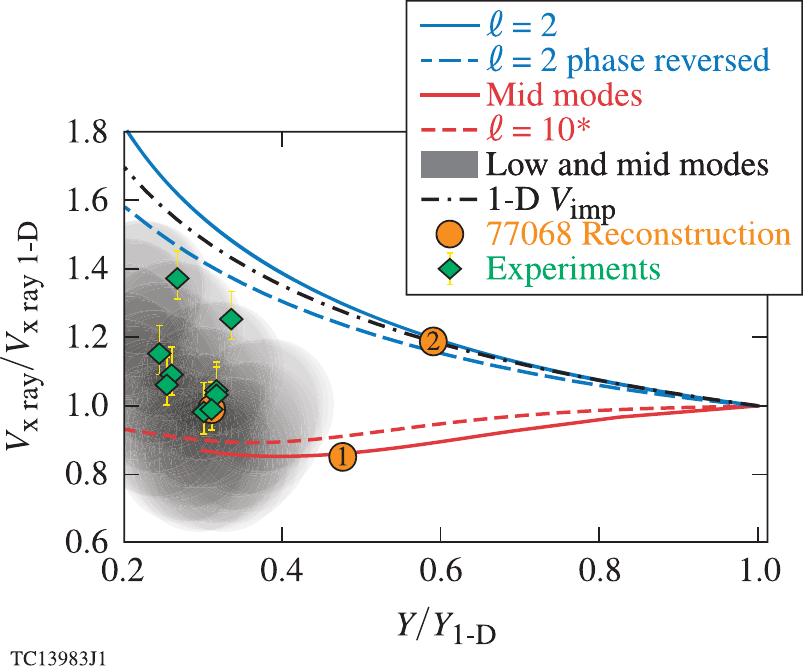}
\caption{\label{fig:Exp_volume} Plot showing the volume of the hot spot, obtained from time-resolved x-ray images and normalized with the 1-D volume ($V_\text{x-ray}/V_\text{x-ray-1D}$), versus the yield degradation $Y/Y_\text{1D}$. The \textit{50 Gbar} shots in Table \ref{tab:Exp_1} are shown in green. The reconstructed shot 77068 is shown in orange (overlapping the x-ray volume for shot 77068), with points (1) and (2) representing the degradation caused by the mid- and low-mode components, separately. The gray-shaded region represents an ensemble of simulations using different amplitude combination of $\ell=2$ and \textit{Mid modes}; it is observed that these reproduce the experiments.}
\end{figure}

The effect of asymmetries on the hot-spot volume is shown in Fig. \ref{fig:Exp_volume}. It is shown that with increasing mode amplitude, the x-ray volume increases for low modes and decreases for mid modes. By cooling the plasma within the RTI bubbles, mid-mode asymmetries cause a reduction in the x-ray--emitting volume. The gray-shaded region (representing the ensemble of simulations) shows that the volume estimated using a combination of low and mid modes is in agreement with the measured volume for the \textit{50 Gbar} shots, illustrating that the experiments can be reconstructed using such combinations of low and mid modes. The effect of an implosion velocity degradation on the x-ray volume has been shown using the dashed black line (1-D V$_\text{imp}$), it follows the scaling $V_\text{x-ray} \sim V_\text{imp}^{-2.14}$. Notice that this curve coincides with the low-mode curves, but it is different from the mid-mode asymmetry curves for the same reasons as previously explained.

\begin{figure}
\includegraphics[width=80mm]{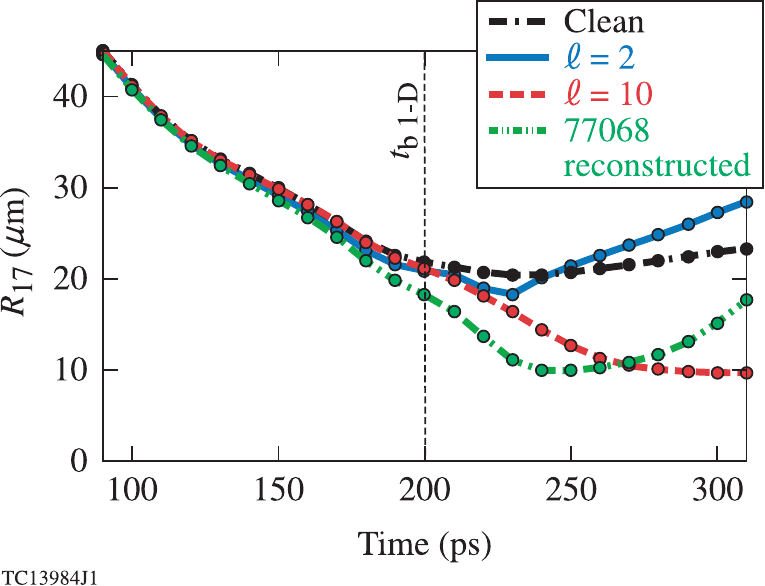}
\caption{\label{fig:Exp_volumeVstime} \ignore{(a) }Plot showing the time evolution of the x-ray $R_{17}$ obtained from simulations. This is shown for the symmetric case (black line), low-mode $\ell=2$ case with $Y/Y_\text{1D}=0.6$ (blue line), mid-mode $\ell=10$ case with $Y/Y_\text{1D}=0.6$ (red line), and the reproduced case with $Y/Y_\text{1D}\approx 0.3$ (green line) for simulations of shot 77068.}
\end{figure}

 The disassembly phase of implosions is different for low- and mid-mode asymmetries, the physical mechanism involved has been discussed in Ref. \onlinecite{Bose_physics_2017}, in this section we discuss signatures in time resolved x-ray images that could aid the detection of mid modes. Time-resolved x-ray images (i.e., with 10 ps gate width) were produced from the simulations using the atomic physics code \textit{Spect3D}.\cite{Exp_Spect3D_1, Exp_Spect3D_2} These images were normalized with the maximum intensity for each image and fit with the following function:
%
%-----------
\begin{equation}
\label{eqn:Exp_supergaussian}
f(x,y) = e^{-\left[ (x/a)^2 + (y/b)^2 \right]^{\eta/2}}.
\end{equation}
%-----------
%
 The $R_{17}$ was obtained from the fit using $R_{17} = \sqrt{a\times b} [-\text{log}(0.17)]^{1/\eta}$. The index $\eta$ represents the index of the super-Gaussian fit, with $\eta=2$ representing a Gaussian function. During the disassembly (i.e., for $t > t_\text{b}$), the $R_{17}$ decreases with time for mid modes, whereas it increases for low modes with respect to the 1-D. A similar trend was also observed for other arbitrary definitions of the radius, i.e. radius at 37$\%$, 50$\%$ and 75$\%$ of peak intensity. \ignore{ In addition, the index $\eta$ decreases for mid-modes and increases for low modes, this is shown in Fig. \ref{fig:Exp_volumeVstime}.} Since detection of mid modes in experiments is challenging, because of the limited spatial resolution of the detectors, the above time-evolution trends in the x-ray images could motivate future experiments.
%
%
%
%%% %%%%%%                                                FIGURE          %%%%%%%%%%%%%%%
\ignore{(b) Plot showing the time evolution of the super-Gaussian $\eta$ for the x-ray emission profile [see Eq. (\ref{eqn:Exp_supergaussian})] for the above-mentioned implosion simulations.}
%
%
%
%\FloatBarrier
%%%%%%%%%%%%%%%%%%%%%%%%%%%%%%%%%%%%%%%%%%
%%%%%%%%%%%%%%%%%%%%%%%%%%%%%%%%%%%%%%%%%%
%%				SECTION2: Shape 
%%%%%%%%%%%%%%%%%%%%%%%%%%%%%%%%%%%%%%%%%%
%%%%%%%%%%%%%%%%%%%%%%%%%%%%%%%%%%%%%%%%%%
\subsection{Shape Analysis of Time-Integrated Self-Emission Images}
\label{sec:Exp_intimage}
%
%
%
%%% %%%%%%                                                FIGURE          %%%%%%%%%%%%%%%
\begin{figure*}
\includegraphics{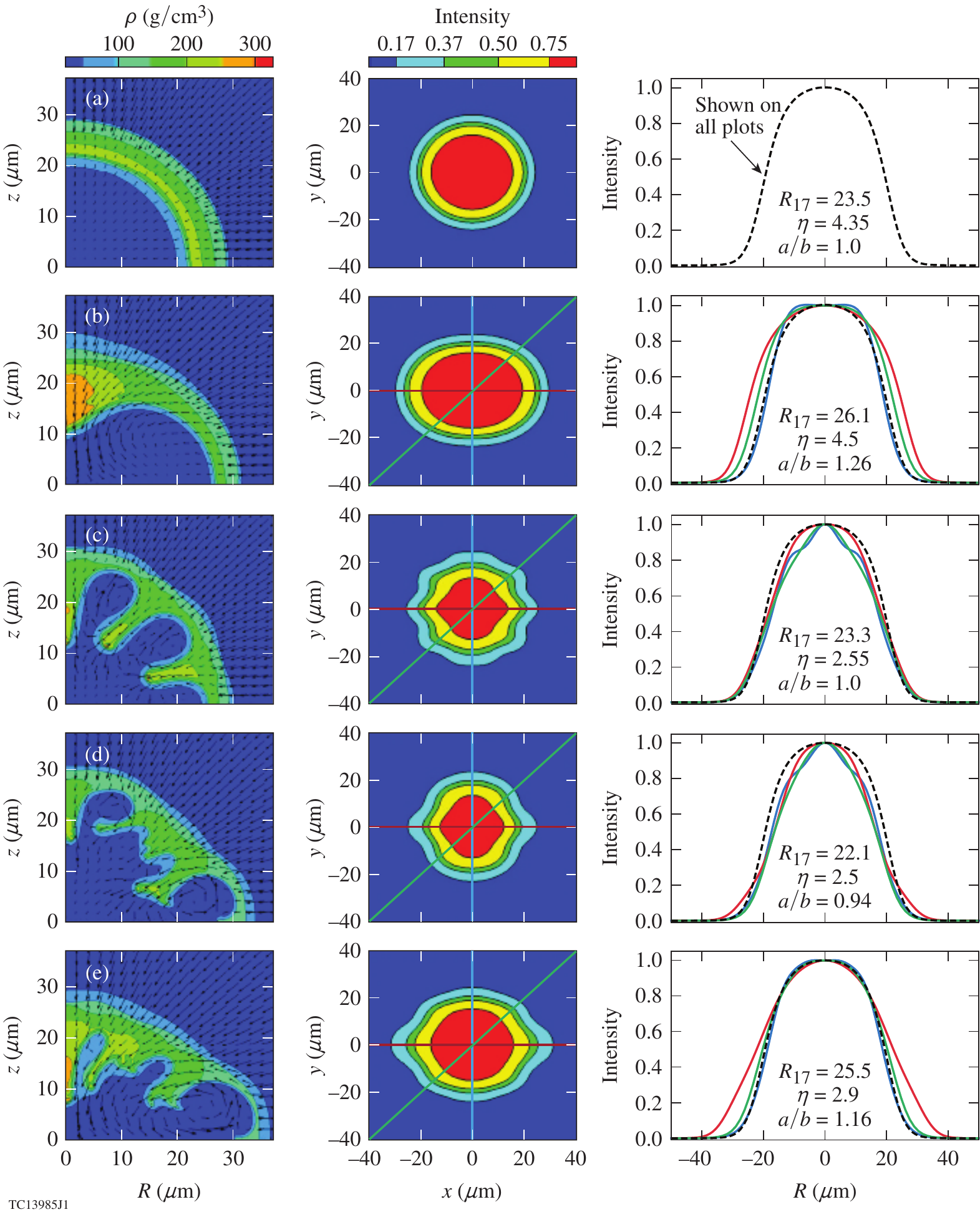}
\caption{\label{fig:Exp_shape1} \footnotesize{Contour plots of the density profile and plasma flow pattern at bang time (first column), time-integrated synthetic x-ray emission images (second column), and image lineouts (third column). The black dashed line represents the lineout of the symmetric image; it is shown on all plots of the third column for reference. The lineouts along the three different axes are labeled with different colors (red, blue, and green). The 2-D super-Gaussian fit parameters have been included. The images for (a) symmetric implosion, (b) $\ell=2$ at $Y/Y_\text{1D}=0.6$, (c) $\ell=10$ at $Y/Y_\text{1D}=0.6$, (d) \textit{Mid modes} (spectrum) with 2$\%$ $\Delta V$ at $Y/Y_\text{1D}=0.47$, and (e) reconstructed shot 77068 are shown.}}
\end{figure*}
%
%
%
%
%
%
%%% %%%%%%                                                TABLE            %%%%%%%%%%%%%%%%
%\clearpage
%\begin{landscape}
\begin{table}
%\begin{center}
\caption{\label{tab:Exp_3} The properties for the time-integrated GMXI\cite{Exp_GMXI} x-ray images from experiments.}
\begin{ruledtabular}

\begin{tabular}{c c c c c}

Shot & $R_{17}$ ($\mu$m) & $\eta$ & $a/b$ & filter  \\

 & $\pm 0.5$ $\mu$m & $\pm$0.2 & $\pm$0.01  & 6.5 mil Be +\\ 

\hline
%\hhline{|=||=|=|=|=|} %\hhline{=|=|=}

\\
78959 &   25.6  & 2.7  & 1.16 &  3 mil Al  \\   %\hline

78963 &  28.1 & 2.3 & 1.17  & 3 mil Al \\  %\hline

78967 & 26.7  & 2.3  & 1.16 & 3 mil Al \\ %\hline

78969 & 27.4  & 2.6   & 1.16  &  3 mil Al \\ %\hline

78971 & 27.1  & 1.9 &  1.20 & 3 mil Al \\ %\hline

77064  & 27.7  & 2.6   & 1.11 &   2 mil Al \\ %\hline

77066  & 26.8  & 2.6  & 1.1 & 2 mil Al \\ %\hline

77068  & 26.7 &  2.69  & 1.16  &  2 mil Al \\ %\hline

77070  & 25.9  & 2.56  &	1.13   & 2 mil Al \\ 
%\hline

\end{tabular}
%\end{center}
\end{ruledtabular}
\end{table}
%\end{landscape}
%%%%%%%%%%%
%
%
%
%
%
%
%
%%% %%%%%%                                                FIGURE          %%%%%%%%%%%%%%%
\begin{figure}
\includegraphics{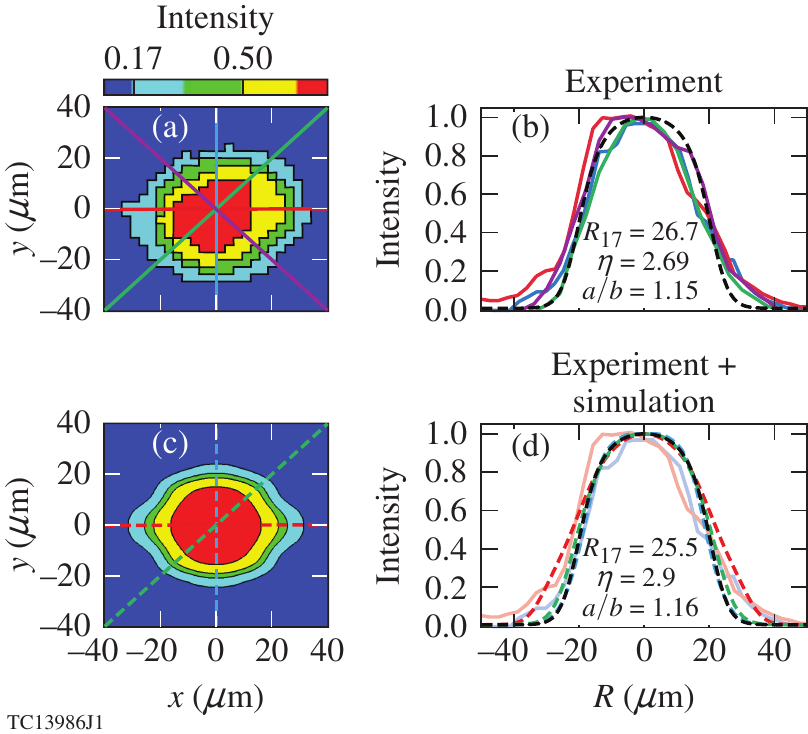}
\caption{\label{fig:Exp_shape2} A comparison between time-integrated x-ray images for shot 77068 obtained from [(a) and (b)] experiments and [(c) and (d)] the reconstructed simulation. The lineouts along the different axes are labeled with different colors (red, blue, green and purple), the lineouts for the experimental image are represented using solid lines [in (b) and (d)], and the simulations are represented using dashed lines [in (d)]. The lineout for the symmetric case is shown with black dashed line [in (b) and (d)] for reference. The super-Gaussian fit parameters for both experiment (b) and simulation (d) are listed.}
\end{figure}

In this section we discuss how asymmetries influence the time-integrated x-ray images. Since the photon statistics (i.e., determined by the number of incident photons) are insufficient for the 10-to-15 ps time-gated images (in Sec. \ref{sec:Exp_gatedimage}), we do not use those images to infer the shape of the hot spot; instead we use the time-integrated images obtained using the gated monochromatic x-ray imaging (GMXI) module.\cite{Exp_GMXI} In Fig. \ref{fig:Exp_shape1} the first column shows the density profile and flow pattern at bang time. The corresponding synthetic self-emission images along with lineouts across a different axis are shown in the second and third columns, respectively. The cross sections were taken through the center of the image; they are marked on the contour plot with the same color as on the intensity plot. The x-ray images were reconstructed with the same filter, point spread function (PSF) and detector response as the experimental shot 77068, i.e., filtered with 6.5 mil of Be and 2 mil of Al, which transmit x rays in the 4-to-8 keV range, and a 7.5 $\mu$m PSF. The images were fit using the function shown in Eq. (\ref{eqn:Exp_supergaussian}). The $R_{17}$ of the time-integrated images, the ellipticity parameter ($a/b$), and the super-Gaussian exponent $\eta$ are calculated from the fit. It is found that low modes cause an increase in the $a/b$ and $R_{17}$, with the index $\eta$ comparable or larger than the 1-D case. In comparison mid modes cause a reduction in the index $\eta$ because the mid modes exhibit several low-temperature bubbles surrounding the hot center, producing a more-gradual intensity variation with radius. The mid modes have negligible effect on the calculated $a/b$ and $R_{17}$.

 Table \ref{tab:Exp_3} shows the properties of the time-integrated x-ray images for the \textit{50 Gbar} shots. It is observed that for all of the shots, the time-integrated $R_{17}$ is larger than the time-resolved images by $\sim$ 3-to-4 $\mu$m (see Table \ref{tab:Exp_1}), this is in consistent agreement with our analysis showing that the time integrated radius ($R_{17}$) is larger than the radius at bang time for low modes ($\ell =2$) in simulations. The $\eta < \eta_\text{1D}$ indicates the presence of mid modes and the $a/b>1$ indicates the presence of low modes in the implosions.

 Figure \ref{fig:Exp_shape2} shows the time-integrated image for shot 77068 and the reconstructed image. The agreement in shape and other parameters ($R_{17}$, $a/b$, and $\eta$) supports the presence of systematic mid modes along with low modes in the \textit{50 Gbar} implosions. In summary, low modes increase the ellipticity parameter ($a/b$) and radius ($R_{17}$) with respect to 1-D from the time-integrated x-ray images, and mid modes produce a lower super-Gaussian index $\eta$. A combination of low- and mid-mode asymmetries can be used to reproduce the experimental images. 
%
%
%
%
%\FloatBarrier
%%%%%%%%%%%%%%%%%%%%%%%%%%%%%%%%%%%%%%%%%%
%%%%%%%%%%%%%%%%%%%%%%%%%%%%%%%%%%%%%%%%%%
%%				SECTION4: Ion Temperature
%%%%%%%%%%%%%%%%%%%%%%%%%%%%%%%%%%%%%%%%%%
%%%%%%%%%%%%%%%%%%%%%%%%%%%%%%%%%%%%%%%%%%
\subsection{Neutron-Averaged Ion Temperature}
\label{sec:Exp_temperature}
%
%
%
%%% %%%%%%                                                FIGURE          %%%%%%%%%%%%%%%
\begin{figure}
\includegraphics[width=80mm]{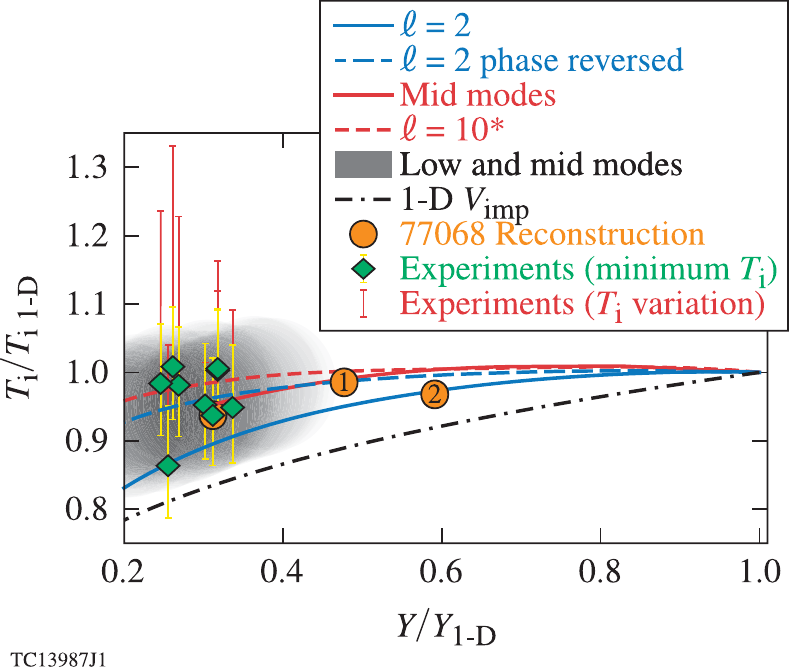}
\caption{\label{fig:Exp_temp} Plot showing degradation in neutron-averaged ion temperature ($T_\text{i}/T_\text{i-1D}$) versus the degradation in yield ($Y/Y_\text{1D}$). The points in green represent the minimum ion temperature measured for the \textit{50 Gbar} shots; the red bar associated with each data point extends to the maximum ion temperature measurement. The reconstructed shot 77068 is shown in orange (overlapping with data); the points (1) and (2) represent degradation caused by the mid-mode and low-mode components separately. The gray-shaded region represents an ensemble of simulations using different amplitude combination of $\ell=2$ and \textit{Mid modes}; it is observed that these reproduce the experiments.}
\end{figure}
%
%
%
%
%
%
%%% %%%%%%                                                FIGURE          %%%%%%%%%%%%%%%
\begin{figure}
\includegraphics[width=80mm]{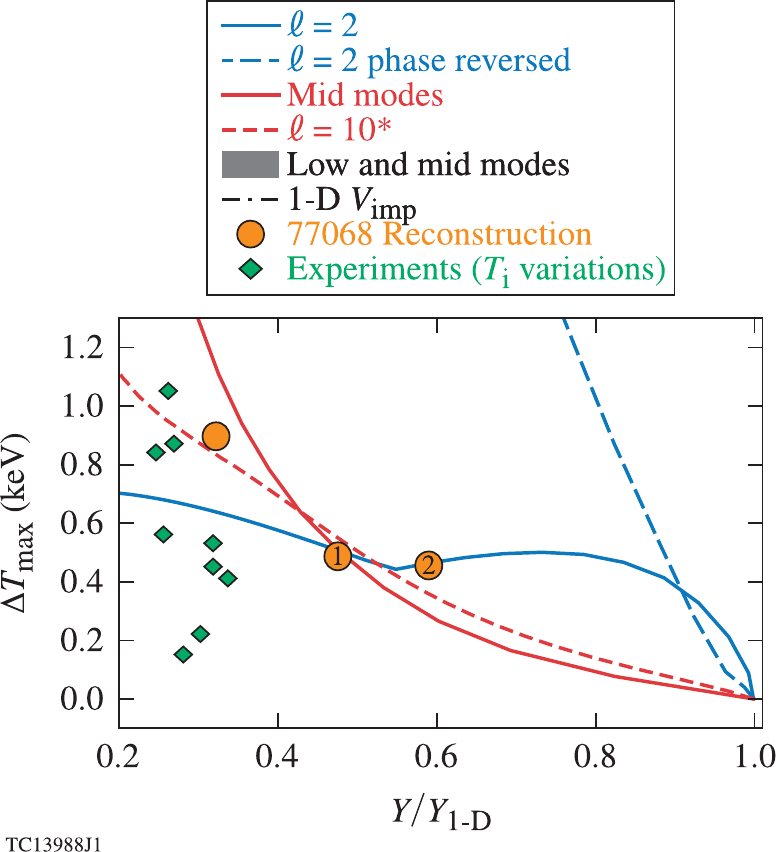}
\caption{\label{fig:Exp_tempVar} 
Plot showing the maximum variation in ion temperature measurements ($\Delta T_\text{max}$) versus degradation in yield ($Y/Y_{1D}$). For the \textit{50 Gbar} experiments, shown in green, the $\Delta T_\text{max}$ is given by $\Delta T_\text{max} = T_\text{i-max}- T_\text{i-min}$ across measurements along different lines of sight. The simulations show maximum variation in ion temperature ($\Delta T_\text{max}$) estimated using Eq. \ref{eqn:Exp_DeltaT}. The reconstructed shot 77068 is shown in orange, with points (1) and (2) representing the degradation caused by the mid- and low-mode components, separately.
}
\end{figure}

Figure \ref{fig:Exp_temp} shows the degradation in ion temperature ($T_\text{i}/T_\text{i-1D}$) with degradation in yield $(Y/Y_\text{1D})$. It is observed that asymmetries cause a small degradation in $T_\text{i}/T_\text{i-1D}$, within $10$-to-$15 \%$ of the 1-D value, for all yield degradation levels above $Y/Y_\text{1D}>0.2$. This is because the temperature of the region of the hot spot that produces fusion neutrons, i.e., the hot region, is only marginally affected by asymmetries (see Ref. \onlinecite{Bose_physics_2017}). Marked in gray are the results from simulations with a combination of low- and mid-mode asymmetries. The points in green, representing the \textit{50 Gbar} experiments, fall within the gray region. In 1-D the temperature scaling with implosion velocity follows $T_\text{i} \sim V_\text{imp}^{0.91}$, which is estimated from the dashed black line. It is observed that at the same yield degradation ($Y/Y_\text{1D}$) level, the temperature is lower for the curve representing implosion velocity degradation (1-D V$_\text{imp}$) as compared to asymmetries.

The variation in ion temperature measurements between detectors is shown using the red bars in Fig. \ref{fig:Exp_temp}; the length of the red bar represents the maximum variation $\Delta T_\text{max}=T_\text{i-Max} -T_\text{i-Min}$ between measurements along different lines of sight for the shot. It is known that flows\cite{brysk_the_1973, chrien_1998, appelbe} in the neutron-producing region of the hot spot, marked with arrows in Fig. \ref{fig:Exp_shape1} (first column), can affect the temperature measurements. This results in a higher apparent temperature, depending on the detector line of sight. The \textit{50 Gbar} implosions exhibit considerable variation in ion temperature measurements. The maximum variation in neutron-averaged ion temperature ($\Delta T_\text{max}$) versus yield degradation level is also shown in Fig. \ref{fig:Exp_tempVar}, the experiments (represented by the points) exhibit shot-to-shot variation in $\Delta T_\text{max}$, this is possibly because of differences in flow effects along different lines of sight. For the simulations, the apparent temperatures (i.e., including flow effects) were calculated using the Murphy\cite{murphy_the_2014} formulation (see Eq. 20 of Ref. \onlinecite{murphy_the_2014})
%
%-----------
\begin{eqnarray}
\label{eqn:Exp_neutronVelocity1}
T^{(\text{app})}_\text{sp/bub} [keV]= \nonumber \\
T_\text{i} [keV] + (m_{\alpha}+m_\text{n}) \left< v^2_\text{sp/bub}\right> [keV] 
\\
\text{with} \hspace{5cm} \nonumber
\\
\label{eqn:Exp_neutronVelocity2}
T_\text{i} =  \frac{\int{\int{T \; n_\text{D} n_\text{T} \left< \sigma v \right> \text{d}V} \text{d}t}}{\int{\int{ n_\text{D} n_\text{T} \left< \sigma v \right> \text{d}V} \text{d}t}}
\\
\label{eqn:Exp_neutronVelocity3}
\left< v^2_\text{sp/bub} \right> = \frac{\int{\int{v^2_\text{sp/bub} n_\text{D} n_\text{T} \left< \sigma v \right> \text{d}V} \text{d}t}}{\int{\int{ n_\text{D} n_\text{T} \left< \sigma v \right> \text{d}V} \text{d}t}}
\end{eqnarray}
%-----------
%
for which we estimate (approximately) the neutron averaged flow broadening along the spike or bubble axis using the above Eqs. \ref{eqn:Exp_neutronVelocity1}-\ref{eqn:Exp_neutronVelocity3}. In the simulations (except the $\ell=$2 \textit{phase reversed} case) the spike axis corresponds to the z-axis (represented by subscript `sp') and the bubble axis is the r-axis (represented by subscript `bub'), see Fig. \ref{fig:Exp_LowMid}; see Fig. \ref{fig:Exp_shape1} for velocity flow field. Notice that the apparent temperature $T^{(\text{app})}_{\text{sp/bub}} \geq T_\text{i}$ the neutron average temperature. The maximum variation possible is estimated using the following, 
%
%
%-----------
\begin{equation}
\label{eqn:Exp_DeltaT}
\Delta T_\text{max} = \text{Max}[T^{(\text{app})}_\text{sp}, T^{(\text{app})}_\text{bub}]- T_\text{i},
\end{equation}
%-----------
%
%
where $T^{(\text{app})}_\text{sp}$ [or $T^{(\text{app})}_\text{bub}$] is the apparent temperature measured by a detector sitting on the spike axis [or bubble axis] and $T_\text{i}$ is the neutron-averaged ion temperature calculated without including the flow effects (as expected, the variation in ion temperature is negligible for symmetric implosions). We find that the $\Delta T_\text{max}$ from experiments and the calculated $\Delta T_\text{max}$ are comparable for implosions with $\ell=2$ and mid modes. The phase-reversed low mode ($\ell=2$ \textit{phase reversed}) produces higher variation in apparent temperature than others in the simulations, this is because these implosions are influenced by significant bulk flow motion within the relatively large neutron producing volume.

Our technique which uses a combination of low and mid modes, can be used to consistently reproduce the neutron-averaged temperature measurements and estimate the variation in temperature for the \textit{50 Gbar} experiments. 
%
%
%
%
%
%\FloatBarrier
%%%%%%%%%%%%%%%%%%%%%%%%%%%%%%%%%%%%%%%%%%
%%%%%%%%%%%%%%%%%%%%%%%%%%%%%%%%%%%%%%%%%%
%%				SECTION5: RhoR estimation
%%%%%%%%%%%%%%%%%%%%%%%%%%%%%%%%%%%%%%%%%%
%%%%%%%%%%%%%%%%%%%%%%%%%%%%%%%%%%%%%%%%%%
\subsection{Implosion Areal Density}
\label{sec:Exp_rhoR}
%
%
%
%%% %%%%%%                                                FIGURE          %%%%%%%%%%%%%%%
\begin{figure}
\includegraphics[width=80mm]{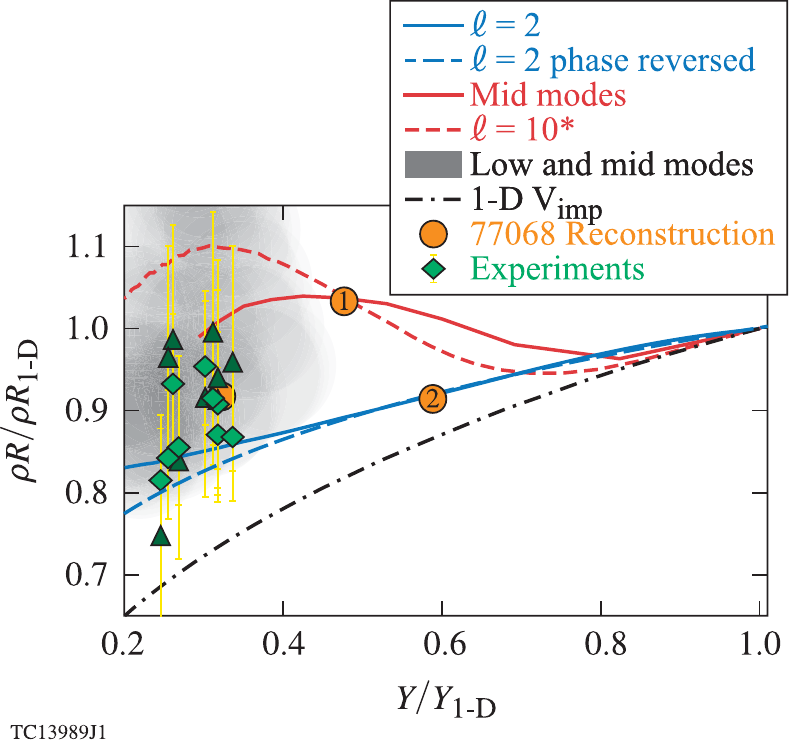}
\caption{\label{fig:Exp_rhoR} Plot showing the degradation in areal density (i.e., $\rho R$ estimated from DSR) versus degradation in yield, the $\rho R$ and yield are normalized with the 1-D estimated values. The NTOF (triangles) and MRS (diamonds) $\rho R$ measurements for the \textit{50 Gbar} shots are shown in green. The reconstructed shot 77068 is shown in orange (overlapping with data), with points (1) and (2) representing degradation caused by the mid-mode and low-mode components, separately. The gray-shaded region represents an ensemble of simulations using different amplitude combination of $\ell=2$ and \textit{Mid modes}; it is observed that these reproduce the experiments.}
\end{figure}
 The effect of asymmetries on the areal density ($\rho R$) is discussed in this section. The $\rho R$'s estimated from the down scattered ratio (DSR) of the neutron spectrum obtained from experiments and simulations are shown in Fig. \ref{fig:Exp_rhoR}. It is observed that the measured $\rho R$'s are comparable to the corresponding 1-D estimated values (from \textit{LILAC}) although the yields are heavily degraded ($Y/Y_\text{1D} \sim 0.3$) in the experiments. In Fig. \ref{fig:Exp_rhoR}, the $\rho R$ scaling with symmetric yield (produced by decreasing the implosion velocity) is shown by the dashed black line (1-D V$_\text{imp}$), it follows $\rho R \sim V_\text{imp}^{1.42}$\ignore{1.68}. In the simulations the $\rho R$s are calculated using the Monte Carlo neutron tracking post-processor code IRIS3D.\cite{IRIS3D} Notice that the $\rho R$ for implosions with asymmetries is always higher than the 1-D V$_\text{imp}$ curve. The $\rho R$ is a parameter dependent on the implosion convergence; for symmetric implosions the yield and $\rho R$ decrease with decreasing convergence according to the 1-D V$_\text{imp}$ curve of Fig. \ref{fig:Exp_rhoR}. Instead, for distorted implosions, the convergence of the spikes can be high, producing a relatively higher $\rho R$, but this does not increase the yield (see Ref. \onlinecite{Bose_physics_2017}). The $\rho R$ for implosions with mid-mode asymmetry (represented by the $\ell=10^*$ and \textit{Mid modes} curves) is comparable to the estimated $\rho R_\text{1D}$. This is because for mid modes, multiple RTI spikes approach the implosion center, producing a compressed plasma with a higher $\rho R$. For the low mode cases ($\ell=2$ and $\ell=2$ \textit{phase reversed}), this effect is relatively small, nevertheless the $\rho R$s at any given $Y/Y_\text{1D}$ are higher than the 1-D $\rho R$ versus yield scaling (represented by the 1-D Vimp). \ignore{reduced because of the multiple modes present in the imposed spectrum. For the $\ell=2$ \textit{phase reversed} case, the spike lies on the implosion waist; the massive spike converges less and the $\rho R$ increase is modest.}

  A combination of low and mid modes (shown by the gray region) could be used to reconstruct the $\rho R$ for the \textit{50 Gbar} shots (shown in green). The measurements along with consideration of the asymmetry trends suggests that a fraction of the measured $\rho R$ is provided by the cold spikes and ablated mass accumulated in the bubbles surrounding the burn volume; therefore, they do not contribute in fusion-yield production but augment the areal density.
%
%
%
%
%\FloatBarrier
%
%%%%%%%%%%%%%%%%%%%%%%%%%%%%%%%%%%%%%%%%%%
%%%%%%%%%%%%%%%%%%%%%%%%%%%%%%%%%%%%%%%%%%
%%				SECTION6: Tau and Burnwidth
%%%%%%%%%%%%%%%%%%%%%%%%%%%%%%%%%%%%%%%%%%
%%%%%%%%%%%%%%%%%%%%%%%%%%%%%%%%%%%%%%%%%%
\subsection{Burnwidth and Bang Time}
\label{sec:Exp_bwandbt}
%
%
%
%
%
%%% %%%%%%                                                FIGURE          %%%%%%%%%%%%%%%
\begin{figure}
\includegraphics[width=80mm]{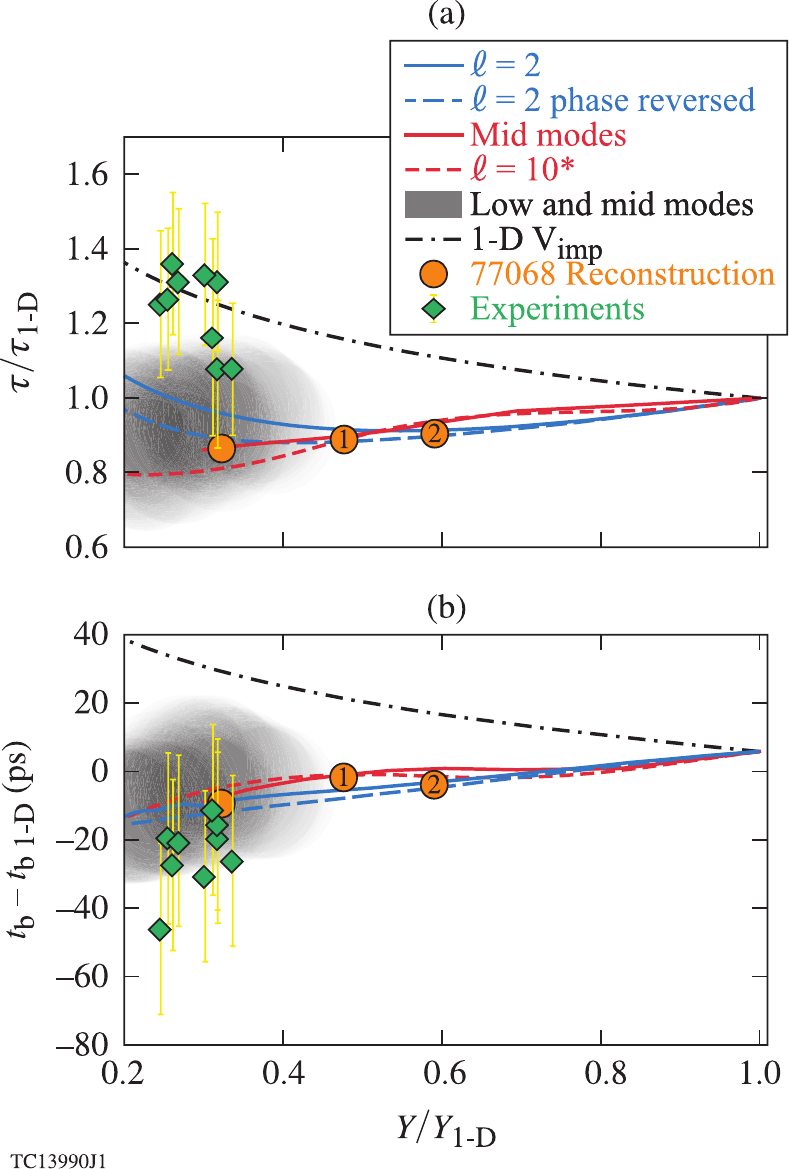}
\caption{\label{fig:Exp_BW_BT}\small{Plots showing (a) burnwidth $( \tau/\tau_\text{1D} )$ and (b) shift in bang time with respect to the 1-D simulations (i.e., $t_\text{b}-t_\text{b-1D}$) versus degradation in yield ($Y/Y_\text{1D}$). The points in green represent the experimental results from the \textit{50 Gbar} implosions (Table \ref{tab:Exp_1}). The reconstructed shot 77068 is shown in orange; the points (1) and (2) represent degradation caused by the mid-mode and low-mode components, separately. The gray-shaded region represents an ensemble of simulations using different amplitude combination of $\ell=2$ and \textit{Mid modes}.}}
\end{figure}
%

%
%
%
%
%%% %%%%%%                                                FIGURE          %%%%%%%%%%%%%%%
%\begin{figure}[h!]
%\begin{center}
%\includegraphics[width=120mm]{asymmetry_exp/BangTime}
%%\includegraphics{fig2_YOCvsIVP_dc}% Here is how to import EPS art
%\caption{\label{fig:Exp_bangtime} Plot showing the variation in bang time with respect to the 1-D simulations (i.e., $t_\text{b}-t_\text{b-1D}$) versus degradation in yield $Y/Y_\text{1D}$.  The points in green represent the \textit{50 Gbar} shots. The reconstructed shot 77068 is shown in orange, with points (1) and (2) representing degradation with the mid-mode and low-mode components separately. The gray shaded region represents an ensemble of simulations using different amplitude combination of $\ell=2$ and mid-modes, it is observed that these reproduce the experiments.
%}
%\end{center}
%\end{figure}
%
%

Figure \ref{fig:Exp_BW_BT}(a) shows a plot of burnwidth degradation ($\tau/\tau_\text{1D}$) with yield degradation ($Y/Y_\text{1D}$). It is observed that the burnwidths from NTD measurements are longer than the 1-D values (from \textit{LILAC}) i.e., $\tau/\tau_\text{1D} > 1$; however, the estimated error in the NTD burnwidths is $\sim \pm 7$ ps. The scaling of burnwidth with implosion velocity is represented using the 1-D V$_\text{imp}$ curve; it follows $\tau \sim V_\text{imp}^{-1.2}$.

 In simulations with asymmetries, the burnwidth shows a modest reduction with degradation in yield. However, for very large low-mode asymmetries (i.e., $Y/Y_\text{1D}<0.4$), the burnwidth increases with decreasing yield, this phenomenon has been described in Ref. \onlinecite{Bose_physics_2017}. A combination of low and mid modes (shown with gray) produce burnwidths that are comparable to the 1-D estimated burnwidth (from \textit{LILAC}) to within $30\%$, but, on average, they are shorter than the burnwidths for the \textit{50 Gbar} experiments.

 Figure \ref{fig:Exp_BW_BT}(b) shows a shift in bang time compared to the 1-D estimated values $(t_\text{b}-t_\text{b-1D})$ with degradation in yield ($Y/Y_\text{1D}$). The bang time from experiments (measured using the NTD) are shifted earlier in time; however, the estimated error in the NTD bang times are considerable ($\approx \pm 25$ ps). Notice that unlike burnwidths, this is in agreement with the asymmetry trends, which also shift the bang time forward; but it is opposite to what an implosion velocity (i.e. 1-D) degradation would do, as shown by the 1-D V$_\text{imp}$ curve for which the bang time occurs later, i.e. $(t_\text{b}-t_\text{b-1D}) > 0$. 

 We propose three possible explanations for the discrepancy between burnwidth and bang time. One possibility is the inaccuracy of the measurements. The NTD measurements for burnwidth and bang time have large error bars and probably are influenced by systematic effects that are not being considered here. It is possible that the actual burnwidths are 10-15 ps shorter, and the actual bang time times are 10-15 ps later than what are measured. The 10-to-15 ps in both burnwidth and bangtime are within the measurement error. This would mean that both are consistent with the trends arising from asymmetries.

 The second possibility is that there is a low-mode asymmetry (not considered here) that is causing an increase in burnwidth, and simultaneously shifts the bang time earlier. We speculate that it could be the outcome of a mode 1 asymmetry. This is based on the observation that for very large $\ell=2$ asymmetry (i.e., $Y/Y_\text{1D}< 0.4$), the $\tau$ increases with a decrease in yield. The implosion dynamics along the orthogonal directions (i.e., along the pole and the waist) are sufficiently mismatched in time, causing a prolonged but inefficient compression, i.e., an increase in $\tau$ although the yield is low. We speculate that this effect on the burnwidth may be enhanced for a $\ell=1$ asymmetry, which involves a larger bulk flow. This could not be verified because the \textit{DEC2D} simulation domain is currently restricted to a $90^{\circ}$ wedge (see Fig. \ref{fig:Exp_LowMid}) instead of a 180$^{\circ}$ wedge required for a mode $1$ simulation.

The most likely explanation is that in addition to a low mode (like the $\ell =2$) and a mid mode (like the $\ell =10$) there is a 1-D degradation in implosion convergence. This would mean that there is a systematic difference in the laser drive that is not accounted for by the laser--plasma coupling models (or equation of state model) in the \textit{LILAC} simulations. Therefore the burnwidths are indeed longer, as measured by the NTD and predicted by the 1-D V$_\text{imp}$ scaling curves. However, the bang time which depends on the history of the acceleration phase, is not correctly captured by the simplistic deceleration-phase scaling (represented by the 1-D $V_\text{imp}$ curves). In experiments, a degradation in implosion convergence can be caused by the following: very short scale nonuniformities arising from laser imprinting or reduced laser-to-capsule drive with respect to simulation, and preheating caused by super-thermal electrons (which decrease the implosion convergence by increasing the implosion adiabat $\alpha$).

%
%
%
%
%
%%%%%%%%%%%%%%%%%%%%%%%%%%%%%%%%%%%%%%%%%%
%%%%%%%%%%%%%%%%%%%%%%%%%%%%%%%%%%%%%%%%%%
%%				SECTION6: Conclusion
%%%%%%%%%%%%%%%%%%%%%%%%%%%%%%%%%%%%%%%%%%
%%%%%%%%%%%%%%%%%%%%%%%%%%%%%%%%%%%%%%%%%%
\section{Conclusions and Future Application}
\label{sec:Exp_summary}
%
%
%%% %%%%%%                                                FIGURE          %%%%%%%%%%%%%%%
%%%%%%%%%%%%%%
\begin{figure}
\includegraphics[width=85mm]{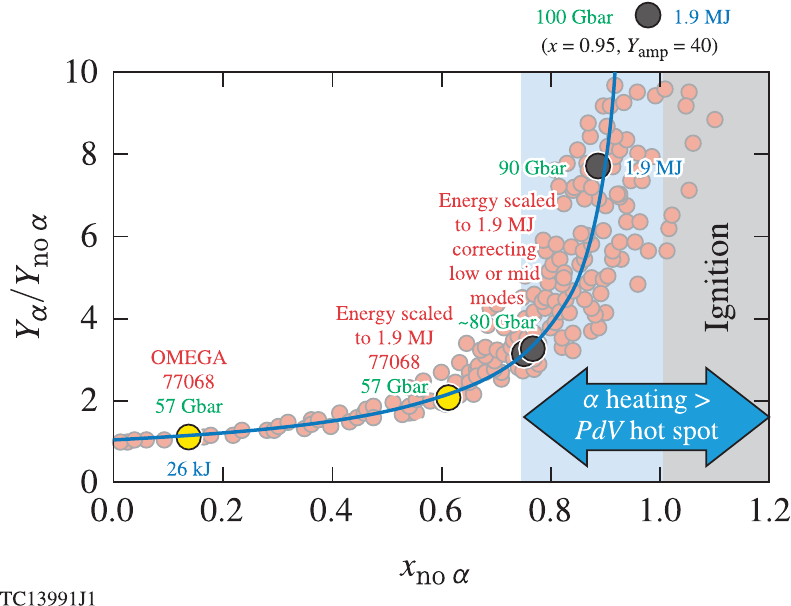}
\caption{\label{Fig3} Plot of yield amplification versus $\chi_{\text{no-}\alpha}$,\cite{Betti-alphaheat} where $\chi_{\text{no-}\alpha}$ is estimated using Eq. 3 of Ref. \onlinecite{Exp_Bose}, 1-D and 2-D simulation results are shown in red and the curve $Y_\alpha/Y_{\text{no-}\alpha} = (1-\chi_{\text{no-}\alpha}/0.96)^{-0.75}$ is shown in blue. The Lawson ignition condition $\chi_{\text{no-}\alpha}\geq 1$ and the burning plasma regime $Q_\alpha \geq 1$ are shown by the gray and blue shaded regions, respectively. OMEGA shot 77068 and its equivalent implosion extrapolated to a 1.9 MJ driver\cite{Exp_Bose} are shown in yellow, they exhibit inferred core pressures of 57 Gbar. Correcting either the low-mode or mid-mode component of this implosion can produce $\approx 80$ Gbar pressure (see Table \ref{tab:Exp_2}), with its performance approaching the burning plasma regime (simulation results are shown in black); improving the asymmetry sources by $\times 0.1$ produces 90 Gbar pressure and the 1-D design has a hot-spot pressure of $\approx$100 Gbar.}
\end{figure}
%%%%%%%%%%%%%%
%
 In this paper a technique to investigate the implosion performance degradation mechanisms was discussed, based on trends in the experimental observables. This was applied to an ensemble of DT cryogenic implosions on OMEGA that achieved hot-spot pressures of $\sim$50 Gbar.\cite{Exp_Regan} It was shown that a combination of low- and mid-mode asymmetries could be used to reconstruct the implosion core.\cite{Exp_Bose} In addition to the presence of low modes, which cause a degradation of the stagnation pressure, it was shown that mid-mode asymmetries have a significant impact on the implosion performance. While it is challenging to image mid-mode asymmetries in implosions, this technique can be used to infer the effect of mid modes on the observables. It was shown that mid modes decrease the hot-spot size (i.e., time-resolved x-ray $R_{17}$) and lead to center-peaked time-integrated x-ray images (i.e., a smaller super-Gaussian exponent $\eta$ compared to a symmetric implosion). This occurs because the region of mid-mode bubbles surrounding the hot center introduces a gradual variation in the x-ray intensity. A consistent explanation for the ion-temperature, areal-density, volume, and pressure measurements for the \textit{50 Gbar} shots was described. The possible reasons behind the modest discrepancies between burnwidth and bang time was discussed based on the measurements and the predicted degradation trends. 

  This paper complements the more detailed analysis of asymmetries provided in Ref. \onlinecite{Bose_physics_2017} with analysis of experiments. It was shown in Ref. \onlinecite{Bose_physics_2017}, that the neutron-averaged observables can differ from the hot-spot volume-averaged quantities; the differences, although small for low modes, are more pronounced for mid-mode asymmetries. In other words, the energy distribution at stagnation is similar for both asymmetry types; however, the fusion reaction distribution is different. This paper described an analysis technique which ventures a consistent correlation between all the experimental observables of the implosion core, based on studies of asymmetries and 1-D degradation. 

 The analysis of several repeats of the cryogenic implosion experiments suggests a systematic degradation mechanism affecting the implosions. A combination of low and mid modes was used to reconstruct all the experimental observables pertaining to the core. It was shown that the experimental observables cannot be explained using either low- or mid-mode asymmetries separately, therefore, a combination was necessary for the reconstruction.

 Mitigation of low- and mid-mode asymmetries would both result in an increase of the fusion yield, however, through an increase of the hot-spot pressure (from 56 Gbar to 80 Gbar) for low modes, and by an increase in the burn volume for mid-modes. Figure \ref{Fig3} shows that an improvement in implosion core symmetry by correcting either the systematic mid or low modes, included in the reconstruction of shot 77068 (and other \textit{50 Gbar} shots \cite{Exp_Regan}), can produce a burning plasma (i.e., $Q_\alpha \geq 1 $, see Ref. \onlinecite{Betti-alphaheat}) when extrapolated to a NIF scale implosion core, i.e., an equivalent 1.9 MJ implosion with symmetric direct illumination (see Ref. \onlinecite{Exp_Bose}). Note that the pressure values shown in Fig. \ref{Fig3} are relevant for the targets discussed in this paper and serve only as an approximate gauge, in-fact implosion performance must be estimated using a Lawson type metric like the $\chi_{\text{no-}\alpha}$.

 In the future this analysis technique will be applied to different 1-D implosion designs (i.e., with different implosion adiabat, obtained from optimization of pulse shape and target dimensions), which would enhance the understanding and possibly lead to identification of the degradation sources for OMEGA direct-drive implosions. 
%
%%%%%%%%%%%%%%%%%%%%%%%%%%%
%%%%%%%%%%%%%%%%%%%%%%%%%%%
%
%
%%%%%%%%%%%%%%%%%%%%%%%%%%%%%%%%%%%%%%%%%%
%%%%%%%%%%%%%%%%%%%%%%%%%%%%%%%%%%%%%%%%%%
%%				SECTION2: Acknowledgments
%%%%%%%%%%%%%%%%%%%%%%%%%%%%%%%%%%%%%%%%%%
%%%%%%%%%%%%%%%%%%%%%%%%%%%%%%%%%%%%%%%%%%
%%%%%%%%%%%%%%%%%%%%%%%%%%%%%%%%%%%%%%%%%%
\section{Acknowledgments}
\label{acknowledgments}
 This research has been supported by the U.S. Department of Energy under Cooperative Agreements DE-FC02-04ER54789 (Office of Fusion Energy Sciences) and DE-NA0001944 (National Nuclear Security Administration), the NYSERDA, and by the Lawrence Livermore National Laboratory under subcontract B614207. This report was prepared as an account of work sponsored by an agency of the U.S. Government. Neither the U.S. Government nor any agency thereof, nor any of their employees, makes any warranty, express or implied, or assumes any legal liability or responsibility for the accuracy, completeness, or usefulness of any information, apparatus, product, or process disclosed, or represents that its use would not infringe privately owned rights. Reference herein to any specific commercial product, process, or service by trade name, trademark, manufacturer, or otherwise does not necessarily constitute or imply its endorsement, recommendation, or favoring by the U.S. Government or any agency thereof. The views and opinions of authors expressed herein do not necessarily state or reflect those of the U.S. Government or any agency thereof.
%
%
%
%%%%%%%%%%%%%%%%%%%%%%%%
%\cleardoublepage

%
%
%

\begin{thebibliography}{5}
%
\bibitem{nuckolls_laser_1972}J. Nuckolls, L. Wood, A. Thiessen, and G. Zimmerman, Nature \textbf{239}, 139 (1972).
%
\bibitem{betti_inertial-confinement_2016}R. Betti and O. A. Hurricane, Nat. Phys. \textbf{12}, 435 (2016).
%
\bibitem{bodner_direct-drive_1998}S. E. Bodner, D. G. Colombant, J. H. Gardner, R. H. Lehmberg, S. P. Obenschain, L. Phillips, A. J. Schmitt, J. D. Sethian, R. L. McCrory, W. Seka, C. P. Verdon, J. P. Knauer, B. B. Afeyan, and H. T. Powell, Phys. Plasmas \textbf{5}, 1901 (1998).
%
\bibitem{asymth_craxton_direct-drive_2015}R. S. Craxton, K. S. Anderson, T. R. Boehly, V. N. Goncharov, D. R. Harding, J. P. Knauer, R. L. McCrory, P. W. McKenty, D. D. Meyerhofer, J. F. Myatt, A. J. Schmitt, J. D. Sethian, R. W. Short, S. Skupsky, W. Theobald, W. L. Kruer, K. Tanaka, R. Betti, T. J. B. Collins, J. A. Delettrez, S. X. Hu, J. A. Marozas, A. V. Maximov, D. T. Michel, P. B. Radha, S. P. Regan, T. C. Sangster, W. Seka, A. A. Solodov, J. M. Soures, C. Stoeckl, and J. D. Zuegel, Phys. Plasmas \textbf{22}, 110501 (2015).
%
\bibitem{lindl_development_1995}J. D. Lindl, Phys. Plasmas \textbf{2}, 3933 (1995).
%
\bibitem{betti_deceleration_2002}R. Betti, K. Anderson, V. N. Goncharov, R. L. McCrory, D. D. Meyerhofer, S. Skupsky, and R. P. J. Town, Phys. Plasmas \textbf{9}, 2277 (2002).
%
\bibitem{Exp_Regan}S. P. Regan, V. N. Goncharov, I. V. Igumenshchev, T. C. Sangster, R. Betti, A. Bose, T. R. Boehly, M. J. Bonino, E. M. Campbell, D. Cao, T. J. B. Collins, R. S. Craxton, A. K. Davis, J. A. Delettrez, D. H. Edgell, R. Epstein, C. J. Forrest, J. A. Frenje, D. H. Froula, M. Gatu Johnson, V. Yu. Glebov, D. R. Harding, M. Hohenberger, S. X. Hu, D. Jacobs-Perkins, R. T. Janezic, M. Karasik, R. L. Keck, J. H. Kelly, T. J. Kessler, J. P. Knauer, T. Z. Kosc, S. J. Loucks, J. A. Marozas, F. J. Marshall, R. L. McCrory, P. W. McKenty, D. D. Meyerhofer, D. T. Michel, J. F. Myatt, S. P. Obenschain, R. D. Petrasso, R. B. Radha, B. Rice, M. Rosenberg, A. J. Schmitt, M. J. Schmitt, W. Seka, W. T. Shmayda, M. J. Shoup III, A. Shvydky, S. Skupsky, A. A. Solodov, C. Stoeckl, W. Theobald, J. Ulreich, M. D. Wittman, K. M. Woo, B. Yaakobi, and J. D. Zuegel, Phys. Rev. Lett. \textbf{117}, 025001 (2016).

\bibitem{Bose_physics_2017}A. Bose, R. Betti, D. Shvarts, and K. M. Woo, Phys. Plasmas \textbf{24}, 102704 (2017).

\bibitem{Exp_LILAC1}J. Delettrez, R. Epstein, M. C. Richardson, P. A. Jaanimagi, and B. L. Henke, Phys. Rev. A \textbf{36}, 3926 (1987).

\bibitem{Exp_LILAC2}I. V. Igumenshchev, W. Seka, D. H. Edgell, D. T. Michel, D. H. Froula, V. N. Goncharov, R. S. Craxton, L. Divol, R. Epstein, R. Follett, J. H. Kelly, T. Z. Kosc, A. V. Maximov, R. L. McCrory, D. D. Meyerhofer, P. Michel, J. F. Myatt, T. C. Sangster, A. Shvydky, S. Skupsky, and C. Stoeckl, Phys. Plasmas \textbf{19}, 056314 (2012).

\bibitem{Exp_LILAC3}V. N. Goncharov, T. C. Sangster, P. B. Radha, R. Betti, T. R. Boehly, T. J. B. Collins, R. S. Craxton, J. A. Delettrez, R. Epstein, V. Yu. Glebov, S. X. Hu, I. V. Igumenshchev, J. P. Knauer, S. J. Loucks, J. A. Marozas, F. J. Marshall, R. L. McCrory, P. W. McKenty, D. D. Meyerhofer, S. P. Regan, W. Seka, S. Skupsky, V. A. Smalyuk, J. M. Soures, C. Stoeckl, D. Shvarts, J. A. Frenje, R. D. Petrasso, C. K. Li, F. Séguin, W. Manheimer, and D. G. Colombant, Phys. Plasmas \textbf{15}, 056310 (2008).

\bibitem{Exp_hurricane}O. A. Hurricane,	D. A. Callahan,	D. T. Casey,	P. M. Celliers,	C. Cerjan,	E. L. Dewald,	T. R. Dittrich,	T. Döppner,	D. E. Hinkel,	L. F. Berzak Hopkins,	J. L. Kline,	S. Le Pape,	T. Ma,	A. G. MacPhee,	J. L. Milovich,	A. Pak,	H.-S. Park,	P. K. Patel,	B. A. Remington,	J. D. Salmonson, P. T. Springer, R. Tommasini, Nature \textbf{506}, 343–348 (2014).

\bibitem{Exp_doppner}T. D$\ddot{\text{o}}$ppner, D.A. Callahan, O.A. Hurricane, D.E. Hinkel, T. Ma, H.-S. Park, L.F. Berzak Hopkins, D.T. Casey, P. Celliers, E.L. Dewald \textit{et al.}, Phys. Rev. Lett. \textbf{115}, 055001 (2015).

\bibitem{Exp_high_foot}O. A. Hurricane, D. A. Callahan, D. T. Casey, E. L. Dewald, T. R. Dittrich, T. Döppner, M. A. Barrios Garcia, D. E. Hinkel, L. F. Berzak Hopkins, P. Kervin, J. L. Kline, S. Le Pape, T. Ma, A. G. MacPhee, J. L. Milovich, J. Moody, A. E. Pak, P. K. Patel, H.-S. Park, B. A. Remington, H. F. Robey, J. D. Salmonson, P. T. Springer, R. Tommasini, L. R. Benedetti, J. A. Caggiano, P. Celliers, C. Cerjan, R. Dylla-Spears, D. Edgell, M. J. Edwards, D. Fittinghoff, G. P. Grim, N. Guler, N. Izumi, J. A. Frenje, M. Gatu Johnson, S. Haan, R. Hatarik, H. Herrmann, S. Khan, J. Knauer, B. J. Kozioziemski, A. L. Kritcher, G. Kyrala, S. A. Maclaren, F. E. Merrill, P. Michel, J. Ralph, J. S. Ross, J. R. Rygg, M. B. Schneider, B. K. Spears, K. Widmann, and C. B. Yeamans, Phys. Plasmas \textbf{21}, 056314 (2014)

\bibitem{Bose_RTIscaling_2015}A. Bose, K. M. Woo, R. Nora, and R. Betti, Phys. Plasmas \textbf{22}, 072702 (2015).

\bibitem{Exp_Bose}A. Bose, K. M. Woo, R. Betti, E. M. Campbell, D. Mangino, A. R. Christopherson, R. L. McCrory, R. Nora, S. P. Regan, V. N. Goncharov, T. C. Sangster, C. J. Forrest, J. Frenje, M. Gatu Johnson, V. Yu. Glebov, J. P. Knauer, F. J. Marshall, C. Stoeckl, and W. Theobald, Phys. Rev. E \textbf{94}, 011201(R) (2016).

\bibitem{Exp_Fred}F. J. Marshall, R. E. Bahr, V. N. Goncharov, V. Yu. Glebov, B. Peng, S. P. Regan, T. C. Sangster, C. Stoeckl, Rev. Sci. Instrum. \textbf{88}, 093702 (2017).

\bibitem{Exp_CVD}G. J. Schmid, R. L. Griffith, N. Izumi, J. A. Koch, R. A. Lerche, M. J. Moran, T. W. Phillips, R. E. Turner, V. Yu. Glebov, T. C. Sangster, and C. Stoeckl, Rev. Sci. Instrum. \textbf{74}, 1828 (2003).

\bibitem{Exp_NTOF1}V. Yu. Glebov, D. D. Meyerhofer, C. Stoeckl, and J. D. Zuegel, Rev. Sci. Instrum. \textbf{72}, 824 (2001).

\bibitem{Exp_NTOF2}V. Yu. Glebov, C. J. Forrest, K. L. Marshall, M. Romanofsky, T. C. Sangster, M. J. Shoup III, and C. Stoeckl, Rev. Sci. Instrum. \textbf{85}, 11E102 (2014).

\bibitem{Exp_MRS}J. A. Frenje, D. T. Casey, C. K. Li, J. R. Rygg, F. H. Séguin, R. D. Petrasso, V. Yu. Glebov, D. D. Meyerhofer, T. C. Sangster, S. Hatchett, S. Haan, C. Cerjan, O. Landen, M. Moran, P. Song, D. C. Wilson, and R. J. Leeper, Rev. Sci. Instrum. \textbf{79}, 10E502 (2008).

\bibitem{Exp_NTD}C. Stoeckl, R. Boni, F. Ehrne, C. J. Forrest, V. Yu. Glebov, J. Katz, D. J. Lonobile, J. Magoon, S. P. Regan, M. J. Shoup III, A. Sorce, C. Sorce, T. C. Sangster, and D. Weiner, Rev. Sci. Instrum. \textbf{87}, 053501 (2016).


\bibitem{Exp_Hu}S. X. Hu, P. B. Radha, J. A. Marozas, R. Betti, T. J. B. Collins, R. S. Craxton, J. A. Delettrez, D. H. Edgell, R. Epstein, V. N. Goncharov, I. V. Igumenshchev, F. J. Marshall, R. L. McCrory, D. D. Meyerhofer, S. P. Regan, T. C. Sangster, S. Skupsky, V. A. Smalyuk, Y. Elbaz, and D. Shvarts, Phys. Plasmas \textbf{16}, 112706 (2009).


\bibitem{Exp_Goncharov}V. N. Goncharov, T. C. Sangster, R. Betti, T. R. Boehly, M. J. Bonino, T. J. B. Collins, R. S. Craxton, J. A. Delettrez, D. H. Edgell, R. Epstein, R. K. Follet, C. J. Forrest, D. H. Froula, V. Yu. Glebov, D. R. Harding, R. J. Henchen, S. X. Hu, I. V. Igumenshchev, R. Janezic, J. H. Kelly, T. J. Kessler, T. Z. Kosc, S. J. Loucks, J. A. Marozas, F. J. Marshall, A. V. Maximov, R. L. McCrory, P. W. McKenty, D. D. Meyerhofer, D. T. Michel, J. F. Myatt, R. Nora, P. B. Radha, S. P. Regan, W. Seka, W. T. Shmayda, R. W. Short, A. Shvydky, S. Skupsky, C. Stoeckl, B. Yaakobi, J. A. Frenje, M. Gatu-Johnson, R. D. Petrasso, and D. T. Casey, Phys. Plasmas \textbf{21}, 056315 (2014).

\bibitem{Exp_Igor}I. V. Igumenshchev, D. T. Michel, R. C. Shah, E. M. Campbell, R. Epstein, C. J. Forrest, V. Yu. Glebov, V. N. Goncharov, J. P. Knauer, F. J. Marshall, R. L. McCrory, S. P. Regan, T. C. Sangster, C. Stoeckl, A. J. Schmitt, and S. Obenschain, Phys. Plasmas \textbf{24}, 056307 (2017).

\bibitem{Exp_Zhou}C. D. Zhou and R. Betti, Phys. Plasmas \textbf{14}, 072703 (2007).

\bibitem{Exp_Edgell}D. H. Edgell, R. K. Follett, I. V. Igumenshchev, J. F. Myatt, J. G. Shaw, and D. H. Froula, Phys. Plasmas \textbf{24}, 062706 (2017); D. H. Edgell, \textit{et al}., \textit{LLE Review Quarterly Report} \textbf{150}, 61 (2017), Laboratory for Laser Energetics, University of Rochester, Rochester, NY, LLE Document No. DOE/NA/1944-1331 (unpublished).

\bibitem{Betti-alphaheat}R. Betti, A. R. Christopherson, B. K. Spears, R. Nora, A. Bose, J. Howard, K. M. Woo, M. J. Edwards, and J. Sanz, Phys. Rev. Lett. \textbf{114}, 255003 (2015).

\bibitem{Exp_SC}C. Cerjan, P. T. Springer, and S. M. Sepke, Phys. Plasmas \textbf{20}, 056319 (2013).

\bibitem{Exp_Bosch}H. S. Bosch and G. M. Hale, Nucl. Fusion \textbf{32}, 611 (1992); \textbf{33}, 1919(E) (1993).

\bibitem{Exp_Spect3D_1}J. MacFarlane, I. Golovkin, P. Wang, P. Woodruff, N. Pereyra, High Energy Density Phys. \textbf{3}, 181-190 (2007)

\bibitem{Exp_Spect3D_2} R. Epstein, S. P. Regan, B. A. Hammel, L. J. Suter, H. A. Scott, M. A. Barrios, D. K. Bradley, D. A. Callahan, C. Cerjan, G. W. Collins, S. N. Dixit, T. Döppner, M. J. Edwards, D. R. Farley, K. B. Fournier, S. Glenn, S. H. Glenzer, I. E. Golovkin, A. Hamza, D. G. Hicks, N. Izumi, O. S. Jones, M. H. Key, J. D. Kilkenny, J. L. Kline, G. A. Kyrala, O. L. Landen, T. Ma, J. J. MacFarlane, A. J. Mackinnon, R. C. Mancini, R. L. McCrory, D. D. Meyerhofer, N. B. Meezan, A. Nikroo, H.-S. Park, P. K. Patel, J. E. Ralph, B. A. Remington, T. C. Sangster, V. A. Smalyuk, P. T. Springer, R. P. J. Town, and J. L. Tucker, AIP Conference Proceedings \textbf{1811}, 190004 (2017)

\bibitem{Exp_GMXI}F. J. Marshall and J. A. Oertel, Rev. Sci. Instrum. \textbf{68}, 735 (1997).

\bibitem{brysk_the_1973}H. Brysk, Phys. Plasmas \textbf{15}, 611 (1973).

\bibitem{chrien_1998}R. E. Chrien, N. M. Hoffman, J. D. Colvin, C. J. Keane, O. L. Landen, and B. A. Hammel, Phys. Plasmas \textbf{5}, 768 (1998).

\bibitem{appelbe}B. Appelbe and J. Chittenden, Plasma Phys. Controlled Fusion \textbf{53}, 045002 (2011).

\bibitem{murphy_the_2014}T. J. Murphy, Phys. Plasmas \textbf{21}, 072701 (2014).

\bibitem{IRIS3D} F. Weilacher, P. B. Radha, \textit{et al.}, \textit{LLE Review Quarterly Report} \textbf{150}, 61 (2017), Laboratory for Laser Energetics, University of Rochester, Rochester, NY, LLE Document No. DOE/NA/1944-1331 (unpublished).
\end{thebibliography}
\end{document}